\documentclass[usegraphicx,usenatbib,useAMS]{mn2e}

\usepackage{multirow}
\usepackage{xcolor}
\usepackage{aas_macros}
\usepackage{amssymb}
\usepackage{xspace}

\newcommand{\gsim}{\gtrsim} 
\newcommand{\lsim}{\lesssim} 

\newcommand{\kms}{\,{\rm km}\,{\rm s}^{-1}}
\newcommand{\Msol}{\,{\rm M}_{\odot}}
\newcommand{\Zsol}{Z_\odot}
\newcommand{\Mpc}{{\rm Mpc}}

\newcommand{\Vc}{V_{\rm c}}
\newcommand{\Vvir}{V_{\rm vir}}
\newcommand{\Vthre}{V_{\rm thresh}}
\newcommand{\Vcrit}{V_{\rm crit}}
\newcommand{\zcrit}{z_{\rm crit}}
\newcommand{\VSN}{V_{\rm SN}}
\newcommand{\gammaSN}{\gamma_{\rm SN}}
\newcommand{\VpSN}{V'_{\rm SN}}
\newcommand{\gammapSN}{\gamma'_{\rm SN}}

\newcommand{\GALFORM}{\textsc{galform}\xspace}
\newcommand{\Planck}{{\em Planck}\xspace}

\title[Constraining SN feedback]
{Constraining SN feedback: a tug of war between reionization and the
  Milky Way satellites}

\author[Hou et al.]  {Jun Hou\thanks{E-mail:
    jun.hou@durham.ac.uk},$^1$ Carlos. S. Frenk,$^1$ Cedric
  G. Lacey,$^1$ Sownak Bose$^1$
  \\
  $^{1}$Institute for Computational Cosmology, Department of Physics,
  University of Durham, South Road, Durham, DH1 3LE, UK}

\begin{document}

\maketitle

\begin{abstract}
  Theoretical models of galaxy formation based on the cold dark matter 
  cosmogony typically require strong
  feedback from supernova (SN) explosions in order to reproduce the
  Milky Way satellite galaxy luminosity function and the faint end of
  the field galaxy luminosity function. However, too strong a SN
  feedback also leads to the universe reionizing too
  late, and the metallicities of Milky Way satellites being too
  low. The combination of these four observations therefore places tight
  constraints on SN feedback. We investigate these
  constraints using the semi-analytical galaxy formation model
  \GALFORM. We find that these observations favour a SN feedback model
  in which the feedback strength evolves with redshift.
  We find that, for our best fit model, half of the
  ionizing photons are emitted by galaxies with rest-frame far-UV
  absolute magnitudes $M_{\rm AB}(1500{\rm \AA})<-17.5$, which implies
  that already observed galaxy populations contribute about half of
  the photons responsible for reionization.
  The $z=0$ descendants of
  these galaxies are mainly galaxies with stellar mass
  $M_*>10^{10}\Msol$ and preferentially inhabit halos with mass $M_{\rm
    halo}>10^{13}\Msol$.
\end{abstract}



\section{introduction}
\label{sec:intro}

Supernova feedback (SN feedback hereafter) is a very important
physical process for regulating the star formation in galaxies
\citep{sn_feedback_larson, sn_feedback_dekel, sn_feedback_wf}. Despite
its importance, SN feedback is not well understood. Perhaps
the best way to improve our understanding of this process is by
investigating its physical properties using hydrodynamical
simulations. This, however, is very difficult to achieve with current
computational power: cosmological hydrodynamical simulations (e.g.\
\citealp{ezw_simulation, illustris_simulation, eagle_simulation}) can
provide large galaxy samples and can follow galaxy evolution spanning
the history of the Universe, but do not have high enough resolution to
follow individual star forming regions, which is needed to understand
the details of SN feedback; conversely high resolution hydrodynamical
simulations \citep[e.g.][]{high_res_feedback2,high_res_feedback3}
can resolve many more details of individual star forming regions, but
do not provide a large sample and cannot follow a long period of
evolution. Because of these limitations, it is worth trying to improve
our understanding of SN feedback in alternative ways. One promising
approach is to extract constraints on SN feedback from theoretical
models of galaxy formation combined with observational constraints.

Among all relevant observations, a combination of four observables 
may be particularly effective because they constrain the strength 
of feedback in opposite directions. These
are the abundance of faint galaxies, including both the faint ends of
the $z=0$ field galaxy luminosity function (hereafter field LF) and
the Milky Way satellite luminosity function (hereafter MW sat LF), the
Milky Way satellite stellar metallicity vs. stellar mass correlation
(hereafter MW sat $Z_*\--M_*$ correlation) and the redshift, $z_{\rm
  re,half}$, at which the Universe was 50\% reionized. The observed
abundance of faint galaxies is very low compared to the abundance of
low mass dark matter halos in the standard cold dark matter (CDM) 
model of cosmogony \citep[e.g.][]{Benson03,missing_satellite1,
missing_satellite2}, which cannot be reproduced by very weak SN
feedback, and this puts a lower limit on the SN feedback strength. On
the other hand, $z_{\rm re,half}$ and the MW sat $Z_*\--M_*$
correlation put upper limits on the SN feedback strength, because too
strong a SN feedback would cause too strong a metal loss and a suppression
of star formation in galaxies, thus leading to too low $Z_*$ at a
given $M_*$, and too low $z_{\rm re,half}$. Also note that this
combination of observations constrains SN feedback over a wide range of
galaxy types and redshifts: the field LF mainly provides constraints
on SN feedback in larger galaxies, with circular velocity $\Vc \gsim
80\kms$, while $z_{\rm re,half}$ mainly constrains SN feedback at
$z\gsim 8$, and the Milky Way satellite observations (MW sat LF and MW
sat $Z_*\--M_*$ correlation) provide constraints on the SN feedback in
very small galaxies, i.e.\ $\Vc\lsim 40\kms$, and probably over a wide
redshift range, from very high redshift to $z \sim 1$. (This is
because recent observations \citep[e.g.][]{de_Boer_fornax,
  dwarf_alpha} indicate that the Milky Way satellites have diverse
star formation histories, with some of them forming all of their stars
very early, and others having very extended star formation histories.)

In this work, we investigate the constraints placed by this
combination of observations on SN feedback using the
semi-analytical galaxy formation model \GALFORM
\citep{cole2000_galform, baugh2005_galform,
  bower2006_galform,Lacey16_model}. A semi-analytical
galaxy formation model is ideal for this aim, because with it one can 
generate large samples of galaxies with high mass resolution,
which is important for simulating both Milky Way satellites and star
formation at high redshift, and it is also computationally feasible to 
explore various physical models and parameterizations.

This paper is organized as follows. Section~\ref{sec:methods}
describes the starting point of this work, the \citet{Lacey16_model}
(hereafter Lacey16) \GALFORM model, as well as extensions of
this model and details of the simulation
runs. Section~\ref{sec:results} presents the results from the Lacey16
and modified models. Section~\ref{sec:discussion} discusses the
physical motivation for some of the modifications, and also which
galaxies drive cosmic reionization and what their $z=0$
descendants are. Finally a summary and conclusions are given in
Section~\ref{sec:summary}.


\section{methods} 
\label{sec:methods}

\subsection{Starting point: Lacey16 model} 
The basic model used in this work is the Lacey16 
\citep{Lacey16_model} model, a recent version of \GALFORM. This
model, and the variants of it that we consider in this paper, all
assume a flat $\Lambda$CDM universe with cosmological parameters based
on the WMAP-7 data \citep{Komatsu11}: $\Omega_{\rm m0}=0.272$,
$\Omega_{\rm v0}=0.728$, $\Omega_{\rm b0}=0.0455$ and
$H_0=70.4~\kms \Mpc^{-1}$, and an initial power spectrum with slope
$n_{\rm s}=0.967$ and normalization $\sigma_8=0.810$.  The Lacey16
model implements sophisticated modeling of disk star formation, improved
treatments of dynamical friction on satellite galaxies and of
starbursts triggered by disk instabilities and an improved stellar
population synthesis model; it reproduces a wide
range of observations, including field galaxy luminosity functions
from $z=0$ to $z=3$, galaxy morphological types at $z=0$, and the
number counts and redshift distribution of submillimetre galaxies. An
important feature of this model is that it assumes a top-heavy IMF for
stars formed in starbursts, which is required to fit the submillimeter data, 
while stars formed by quiescent star
formation in disks have a Solar neighbourhood IMF. Stellar
luminosities of galaxies at different wavelengths, and the production
of heavy elements by supernovae, are predicted self-consistently, 
allowing for the varying IMF.

SN feedback is modeled in this and earlier versions of \GALFORM as
follows. SN feedback ejects gas out of galaxies, and thus reduces the
amount of cold gas in galaxies, regulating the star formation. The
gas ejection rate is formulated as:
\begin{equation} 
\dot{M}_{\rm eject}=\beta \psi,
\label{eq:beta_def}
\end{equation} 
where $\dot{M}_{\rm eject}$ is the mass ejection rate, $\psi$ is the
star formation rate and the mass-loading factor, $\beta$, encodes the
details of SN feedback models. In the approximation of
instantaneous recycling that we use here, in which we neglect the
time delay between the birth and death of a star, the supernova rate,
and hence also the supernova energy injection rate, are proportional
to the instantaneous star formation rate $\psi$.

\begin{figure*}
 \centering
\includegraphics[width=1.0\textwidth]{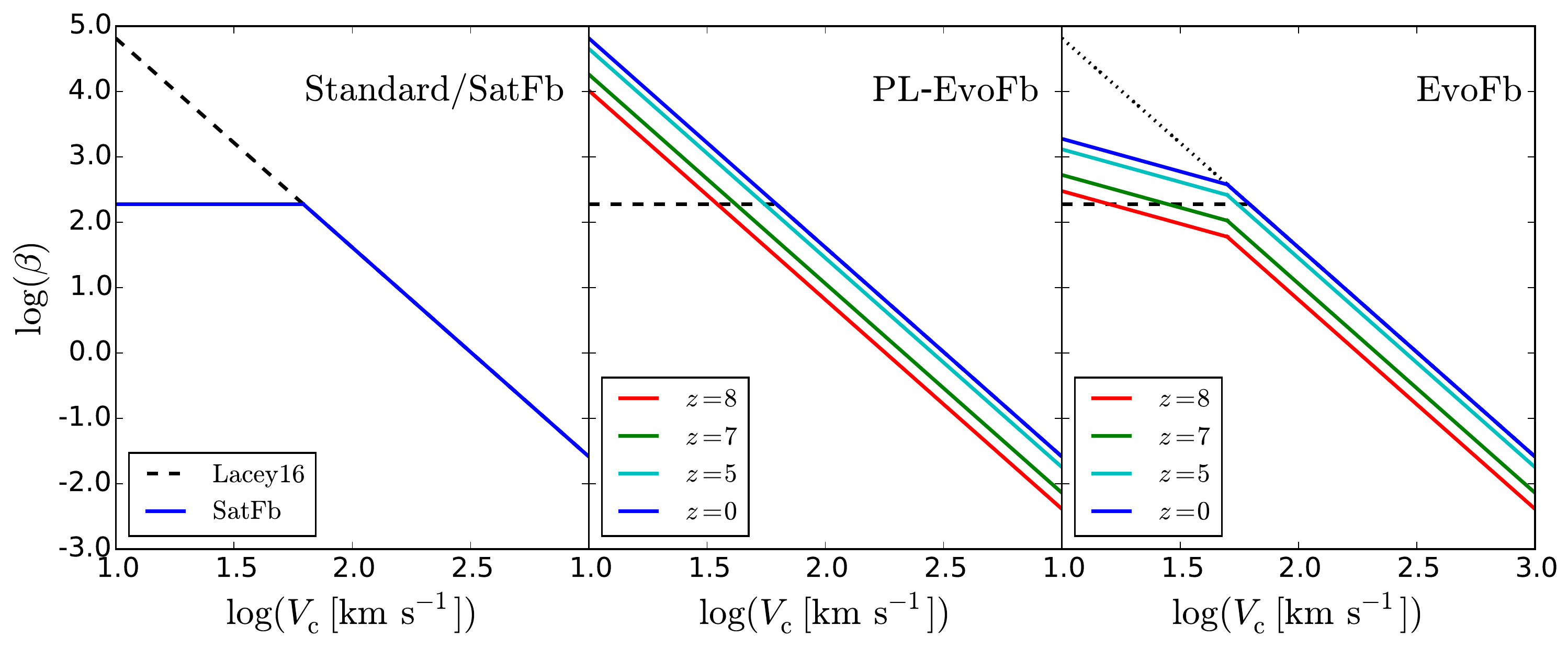}
\caption{Mass-loading factor, $\beta$, as a function of circular
  velocity, $\Vc$, and redshift, $z$, for the different supernova
  feedback models used in this work. {\bf Left panel:} The dashed
  black line shows $\beta$ in the Lacey16 model, while the solid blue
  line shows $\beta$ for the SatFb model. {\bf Middle
      panel:} $\beta$ for the PL-EvoFb model.  Different colours
    indicate different redshifts (from top to bottom, redshift increases 
    from $0$ to $8$). This model is identical to the
    Lacey16 model for $z\leq 4$ (solid blue line). The SatFb model is also
    plotted as a dashed line for reference. {\bf Right panel:} $\beta$
  for the EvoFb model. Different colours indicate different redshifts
  (from top to bottom, redshift increases from $0$ to $8$). 
  $\beta$ for the Lacey16 and SatFb models are also plotted for
  reference, and are shown by the black dotted and dashed lines
  respectively.}

\label{fig:beta}
\end{figure*}

In the Lacey16 model, $\beta$ is set to be a single power law in 
galaxy circular velocity, $\Vc$, specifically,
\begin{equation} 
\beta=\left( \frac{\Vc}{\VSN}
\right)^{-\gammaSN},
\label{eq:single_powerl}
\end{equation} 
where $\VSN$ and $\gammaSN$ are two free parameters. In the Lacey16
model, $\VSN=320\kms$ and $\gammaSN=3.2$. $\beta$ as a
function of $\Vc$ for the Lacey16 model is illustrated in the left
panel of Fig.~\ref{fig:beta}.

As shown in Figs.~\ref{fig:z_re} and \ref{fig:mw_metal}, the above
single power-law SN feedback model is disfavored by the combination of
the four observational constraints mentioned in \S\ref{sec:intro}. We
therefore investigate some modified SN feedback models and test them
against the same set of observations. These modified models are
described next.

\subsection{Modified SN feedback models} 
In the modified SN feedback models we assume a broken power law for
$\beta$, with a change in slope below a circular velocity, $\Vthre$:
\begin{equation} 
\beta =\left\{
\begin{array}{cc} (\Vc/\VSN)^{-\gammaSN} &
\Vc\geq \Vthre \\ (\Vc/\VpSN)^{-\gammapSN} & 
\Vc< \Vthre
\end{array} \right. .
\label{eq:modified_fb}
\end{equation}  
Here $\VSN$, $\gammaSN$, $\Vthre$ and $\gammapSN$ are free
parameters, while $\VpSN$ is fixed by the condition that the two
power laws should join at $\Vc=\Vthre$.

\subsubsection{Saturated feedback model} 
In this class of models we set $\gammapSN<\gammaSN$, so
that the mass-loading factor, $\beta$, for $\Vc < \Vthre$
is lower than in the single power-law model. Note that we require
$\gammapSN\ge 0$, because a negative $\gammapSN$ would
predict an anti-correlation between galaxy stellar metallicity and
stellar mass, in contradiction with observations of the MW satellites.

A similar feedback model, with $\gammapSN=0$, was previously
used by \citet{font2011}, which showed that it improved the agreement of
\GALFORM model predictions with Milky Way observations. However, in
the present work, the observational constraints are more stringent than in
\citet{font2011}, because here not only are Milky Way observations
considered, but also the field LF and the reionization redshift.

In this work, we investigate a specific saturated feedback model, with
$\Vthre=62\kms$ and $\gammapSN=0$, which implies that $\beta$ is a
constant for galaxies with $\Vc<62\kms$ but reduces to the standard
Lacey16 form for $\Vc>\Vthre$. We call this specific saturated
feedback model SatFb. The mass-loading factor for this model is also
illustrated in the left panel of Fig.~\ref{fig:beta}.


\subsubsection{Evolving feedback model} 
This class of model has weaker SN feedback strength
  at high redshift. Here we investigate two specific models. For the
  first one, called PL-EvoFb, the feedback strength is a single power
  law in $\Vc$ at any redshift, but the normalization changes with
  redshift, being identical to the Lacey16 model for $z\leq 4$, but
  lower at high redshifts, Specifically, this model has
  $\gammapSN=\gammaSN=3.2$ (as in Lacey16) and
\begin{equation} 
\VSN\ [{\rm km}\,{\rm s}^{-1}]=\left\{
\begin{array}{cc} 180 & z > 8 \\ -35z+460 & 4 \leq z \leq 8 \\ 320 & z
< 4
\end{array} \right. .
\label{eq:evolving_vhot}
\end{equation}
The general behaviour of this model is motivated by the results of
\citet{lagos_feedback}, who predicted mass-loading factors from a
detailed model of SN-driven superbubbles expanding in the ISM. (The
\citeauthor{lagos_feedback} model was however incomplete, in that it
considered only gas ejection out of the galaxy disk, but not out of
the halo.) The mass loading factor, $\beta$, for this model is
illustrated in the middle panel of Fig.~\ref{fig:beta}.

The second model that we try, called EvoFb, has a
normalization that evolves with redshift as in the PL-EvoFb model,
but also has a shallower $\Vc$-dependence at low
$\Vc$. Specifically, this model has $\gammaSN=3.2$ (as in Lacey16),
$\gammapSN=1.0$, $\Vthre=50\kms$ and $\VSN(z)$ as given in
Eq~(\ref{eq:evolving_vhot}). For $\Vc>50\kms$, this model is
identical to the PL-EvoFb model, but it has weaker feedback for
$\Vc\leq50\kms$. The saturation in $\beta$ at low $\Vc$ is therefore
weaker than in the SatFb model. The mass loading factor for this
model is illustrated in the right panel of Fig.~\ref{fig:beta}.

The physical motivation for introducing the redshift evolution in the
SN feedback will be discussed further in \S\ref{sec:discussion}.


\subsection{The redshift of reionization and photoionization feedback}

\label{sec:reionization_calculation}

We estimate the redshift of reionization predicted by a \GALFORM
model by calculating the ratio, ${\cal R}(z)$, of the number density
of ionizing photons produced up to that redshift to the number density
of hydrogen nuclei:
\begin{equation} 
{\cal R}(z)=\frac{\int_{z}^{\infty} \epsilon (z')
dz'}{n_{\rm H}},
\label{eq:reion1}
\end{equation}  
where $\epsilon (z')$ is the number of hydrogen-ionizing photons
produced per unit comoving volume per unit redshift at redshift $z'$,
and $n_{\rm H}$ is the comoving number density of hydrogen nuclei.

The Universe is assumed to be fully ionized at a redshift, $z_{\rm
  re,full}$, for which,
\begin{equation} 
{\cal R}(z_{\rm re,full})=\frac{1+N_{\rm rec}}{f_{\rm esc}},
\label{eq:zre_full}
\end{equation} 
where $N_{\rm rec}$ is the mean number of recombinations per hydrogen
atom up to reionization, and $f_{\rm esc}$ is the fraction of ionizing
photons that can escape from the galaxies producing them into the
IGM. In this paper we adopt $N_{\rm rec}=0.25$ and $f_{\rm esc}=0.2$,
and thus the threshold for reionization is ${\cal R}(z_{\rm
  re,full})=6.25$. Below we justify these choices. 

Our estimation of the reionization redshift using
  ${\cal R}(z)$ (Eqs~(\ref{eq:reion1}) and (\ref{eq:zre_full}))
  appears to be different from another commonly used estimator based
  on $Q_{\rm HII}$, defined as the volume fraction of ionized
  hydrogen, with reionization being complete when $Q_{\rm HII}=1$, but
  in fact they are essentially equivalent. The evolution equation for
  $Q_{\rm HII}$ is given in \cite{Madau1999_reionization} as
  $\dot{Q}_{\rm HII}=\dot{n}_{\rm ion}/n_{\rm H}-Q_{\rm
    HII}/\bar{t}_{\rm rec}$, where $n_{\rm ion}$ is the comoving
  number density of ionizing photons escaping into IGM and
  $\bar{t}_{\rm rec}$ is the mean recombination time
  scale. Integrating both sides of this equation from $t=0$ to the
  time $t=t_{\rm re,full}$ when reionization completes, one obtains
\begin{eqnarray}
\lefteqn{Q_{\rm HII}(t_{\rm re,full})} \nonumber \\
  & =  & \frac{\int_{0}^{t_{\rm re,full}}\dot{n}_{\rm ion}\, dt}{n_{\rm H}}
      -\frac{\int_{0}^{t_{\rm re,full}}[n_{\rm H}Q_{\rm HII}/\bar{t}_{\rm rec}]\, dt}{n_{\rm H}} \nonumber \\
  & = & \frac{f_{\rm esc}\int_{z_{\rm re,full}}^{\infty}\epsilon (z') dz'}{n_{\rm H}} - \frac{n_{\rm rec,tot}}{n_{\rm H}} 
\end{eqnarray}
where $n_{\rm rec,tot}$ is the mean number of recombinations per
comoving volume up to $z_{\rm re,full}$. Setting $Q_{\rm HII}(t_{\rm
  re,full})=1$ and defining $N_{\rm rec} = {n_{\rm rec,tot}}/{n_{\rm H}}$,
one then obtains Eq~(\ref{eq:zre_full}) for $z_{\rm re,full}$.

With the expression for $\bar{t}_{\rm rec}$ given by \cite{AGN_reionization2}
(their Eqn~4), $N_{\rm rec}$ can be expressed as
\begin{equation}
N_{\rm rec}=\int_{0}^{t_{\rm re,full}}[Q_{\rm HII}(1+\chi)\alpha_{\rm B}(1+z)^3C_{\rm RR}]dt \label{eq:N_rec_madau}
\end{equation}
where $\chi = 0.083$, $\alpha_{\rm B}$ is the case-B recombination
rate coefficient and $C_{\rm RR}$ the clumping factor.  Using the
clumping factor in \cite{clumping_factor_shull} and solving the equation
 for $Q_{\rm HII}$, Eqn~(\ref{eq:N_rec_madau}) gives $N_{\rm rec}$ in
the range $0.13-0.34$ for our four different SN feedback models and an
IGM temperature, $T=1-2\times 10^4\,{\rm K}$. Our choice of $N_{\rm
  rec}=0.25$ lies within this range; note that Eqn~(\ref{eq:zre_full})
is not very sensitive to $N_{\rm rec}$ when its value is much smaller than
$1$. Our choice for $N_{\rm rec}$ is lower than the values assumed in
some previous works \citep[e.g.][]{reion_1} because recent simulations
give lower clumping factors (see \citealt{clumping_factor_finlator} and
references therein).

The calculation of $\epsilon (z')$ requires a knowledge of the
ionizing sources. The traditional assumption has been that these
sources are mainly star-forming galaxies, but recently there have been
some works
\citep[e.g.][]{AGN_reionization1,AGN_reionization2,AGN_reionization3}
suggesting that AGN could be important contributors to reionization of
hydrogen in the IGM. Although AGN might be important for reionization,
these current works rely on extrapolatiing the AGN luminosity function
faintwards of the observed luminosity limit, and also extrapolating
the observations at $z\leq 6$ to $z\sim 10$, in order to obtain a
significant contribution to reionization from AGN. These
extrapolations are uncertain, therefore in this work we ignore any AGN
contribution and assume that the ionizing photon budget for
reionization is dominated by galaxies. We discuss how the AGN
contribution affects our conclusion in more detail in
\S\ref{sec:uncertainty}.

The value of the escape fraction, $f_{\rm esc}$, is also
uncertain. Numerical simulations including gas dynamics and radiative
transfer have given conflicting results: \cite{simulation_f_esc1}
estimated $f_{\rm esc} \sim 0.1$, with $f_{\rm esc}\sim 0.2$ for
starbursts, while \cite{simulation_f_esc2} found much lower
values. These differences between simulations may result from
differences in the modelling of the ISM or in how well it is resolved,
both of which are challenging problems. Observationally, it is
impossible to measure $f_{\rm esc}$ directly for galaxies at the
reionization epoch, because escaping ionizing photons would, in any
case, be absorbed by the partially neutral IGM. Thus, one has to rely
on observations of lower redshift galaxies for clues to its
value. 

Observations of Lyman-break galaxies at $z=3-4$ suggest a relatively
low value, $f_{\rm esc}\sim 0.05$ \citep{Obs_f_esc1}, while
observations of local compact starburst galaxies show indirect
evidence for higher $f_{\rm esc}$ \citep[e.g.][]{Obs_f_esc2};
\cite{Obs_f_esc3} estimated $f_{\rm esc}=0.21$ for one local
example. It is therefore important to determine what class of
currently observed galaxies are the best analogues of galaxies at the
reionization epoch. In our simulations, galaxies at high redshift tend
to be compact and, in addition, the galaxies dominating the ionizing
photon budget are starbursts (see Fig.~\ref{fig:re_f_burst}), so, as
argued by \cite{Sharma_16}, they may well have similar escape
fractions to local compact starburst galaxies. \cite{Sharma_16}
provide further arguments that support our choice of $f_{\rm
  esc}=0.2$. We discuss how the uncertainties in $f_{\rm esc}$ affect
our conclusions in more detail in \S\ref{sec:uncertainty}.

Note that, as advocated by \cite{Sharma_16} we only assume $f_{\rm
  esc}=0.2$ for $z\geq 5$; for lower redshifts, $f_{\rm esc}$ may drop
to low values, consistent with recent studies which argue that $f_{\rm
  esc}$ evolves with redshift and increases sharply for $z>4$
\citep[e.g.][]{reion_threo1,reion_threo2} .

Observations of the CMB directly constrain the electron scattering
optical depth to recombination, which is then converted to a
reionization redshift by assuming a simple model for the redshift
dependence of the ionized fraction. Papers by the {\em WMAP} and {\em
  Planck} collaborations \citep[e.g.][]{planck2013_XVI} typically
express the reionization epoch in terms of the redshift, $z_{\rm
  re,half}$, at which the IGM is 50\% ionized, by using the simple
model for non-instantaneous reionization described in Appendix~B of
\citet{CMB_simple_reion_model}. For comparing with such observational
estimates, we therefore calculate $z_{\rm re,half}$ from \GALFORM by
assuming ${\cal R}(z_{\rm re,half})= \frac{1}{2} {\cal R}(z_{\rm
  re,full})$. For the abovementioned choices of $N_{\rm rec}$ and
$f_{\rm esc}$, this is equivalent to ${\cal R}(z_{\rm re,half})=3.125$.

Reionization may suppress galaxy formation in small halos, an effect
called photoionization feedback \citep{photonionization_feedback1,
photonionization_feedback2,photonionization_feedback3}. 
In this work, the photoionization
feedback is modeled using a simple approximation \citep{Benson03}, in
which dark matter halos with circular velocity at the virial radius
$\Vvir < \Vcrit$ have no gas accretion or gas cooling for
$z<\zcrit$. As shown by \citet{benson_reion} and \citet{font2011},
this method provides a good approximation to a more complex,
self-consistent photoionization feedback model. Here, $\Vcrit$ and
$\zcrit$ are two free parameters. In this paper, unless otherwise
specified, we adopt $\zcrit=z_{\rm re,full}$ and $\Vcrit=
30\kms$. This value of $\Vcrit$ is consistent with the hydrodynamical
simulation results of \citet{okamoto2008}. Note that this method does
not necessarily imply that star formation in galaxies in halos with
$\Vvir < \Vcrit$ is turned off immediately after $z=z_{\rm
  re,full}$. The star formation in these galaxies can continue as long
as the galaxy cold gas reservoir is not empty.

\subsection{Simulation runs} 
Studying reionization requires resolving galaxy formation in low
mass halos ($M_{\rm vir} \sim 10^{8} \-- 10^{10}\,{\rm M}_{\odot}$) at
high redshifts ($z \sim 7 \-- 15$), and thus very high mass
resolution for the dark matter halo merger trees. The easiest way to
achieve this high resolution is to use Monte Carlo (MC) merger trees.

Studying the properties of the Milky Way satellites also requires very
high mass resolution because the
host halos of these small satellites are small. This too is
easily achieved using MC merger trees. Furthermore, because
building MC merger trees is computationally inexpensive, it is
possible to build a large statistical sample of Milky Way-like halos
to study their satellites.

In this work we generate MC merger trees using
the method of \citet{Parkinson08}. To study reionization, we ran
simulations starting at $z_{\rm start}=20$ down to different
final redshifts, $z_{\rm end}$, to derive $\epsilon (z)$ defined in
Eq~(\ref{eq:reion1}) at $z=5\--15$ and the $z=0$ field LF. We scale
the minimum progenitor mass in the merger trees as $(1+z_{\rm
  end})^{-3}$, with a minimum resolved mass, $M_{\rm res}=7\times 10^{9}\Msol$ for $z_{\rm
  end}=0$. We have tested that these choices are sufficient to derive
converged results. For the Milky Way satellite study, the present-day
host halo mass is chosen to be in the range $5\times 10^{11}\-- 2\times
10^{12}\Msol$, which represents the current observational
constraints on the halo mass of the Milky Way, and we sample this
range with five halo masses evenly spaced in log(mass). For each of
these halo masses, \GALFORM is run on $100$ MC merger trees, with
minimum progenitor mass $M_{\rm res}=1.4\times 10^{6}\Msol$, which is
small enough for modeling the Milky Way satellites, and $z_{\rm
  start}=20$ and $z_{\rm end}=0$. We do not attempt to select Milky Way-like host
galaxies, because we found that the satellite properties 
correlate better with the host halo mass than with the host galaxy
properties.


\section{results} 
\label{sec:results}
In this section, we show how the results from the different models
compare with the key observational constraints that we have
identified, namely: the field galaxy luminosity functions at
$z=0$; the redshift of reionization; the MW satellite galaxy luminosity function; and
the stellar metallicity vs stellar mass relation for MW satellites.

\begin{figure*} 
\centering
\includegraphics[width=1.0\textwidth]{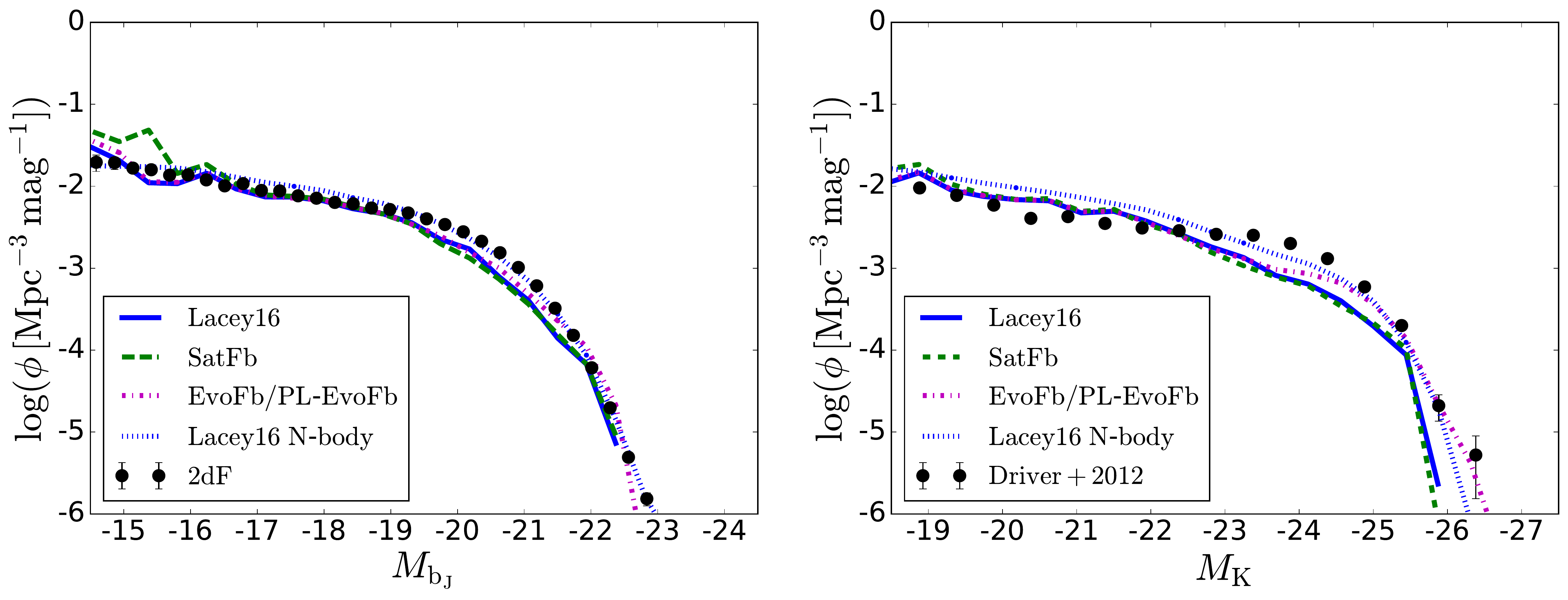}
\caption{$z=0$ field luminosity functions. The left panel shows the
  $b_J$-band luminosity function and the right panel the $K$-band
  luminosity function. In both panels the solid, dashed and
  dashed-dotted lines with different colours show the predictions
  using Monte Carlo merger trees for different SN feedback models, as
  indicated in the line key, while the blue dotted lines are for the
  Lacey16 model run with N-body merger trees. The magenta lines show
  the results of the EvoFb model, but the results for the PL-EvoFb
  model are almost identical. The black points with errorbars are
  observational data, from \citet{bj_lf_2df} for the $b_J$-band and
  from \citet{k_lf_driver} for the $K$-band.}

\label{fig:field_lf}
\end{figure*}

\subsection{Lacey16 model} 
We begin by showing the results for the default Lacey16 model, since
this then motivates considering models with modified SN feedback.
Fig.~\ref{fig:field_lf} shows the $b_J$- and $K$-band field LFs of
different models at $z=0$ (left and right panels respectively). The
dotted blue lines show the LFs calculated using N-body merger trees,
as used in the original \citet{Lacey16_model} paper to calibrate the
model parameters. The fit to the observed LFs is seen to be very
good. The solid blue lines show the predictions with identical model
parameters but instead using MC merger trees, as used in the remainder
of this paper. The run with MC merger trees gives slightly lower LFs
than the run with the N-body trees around the knee of the LF, but at
lower luminosities, the results predicted using MC and N-body merger
trees are in good agreement. We remind the reader that we use MC
merger trees in the main part of this paper in order to achieve the
higher halo mass resolution that we need for the other observational
comparisons. Since the differences in the LFs between the two types of
merger tree are small, and barely affect the faint end of the field
LFs which are the main focus of interest here, we do not consider them
important for this paper.

\begin{figure*} 
\centering
\includegraphics[width=1.0\textwidth]{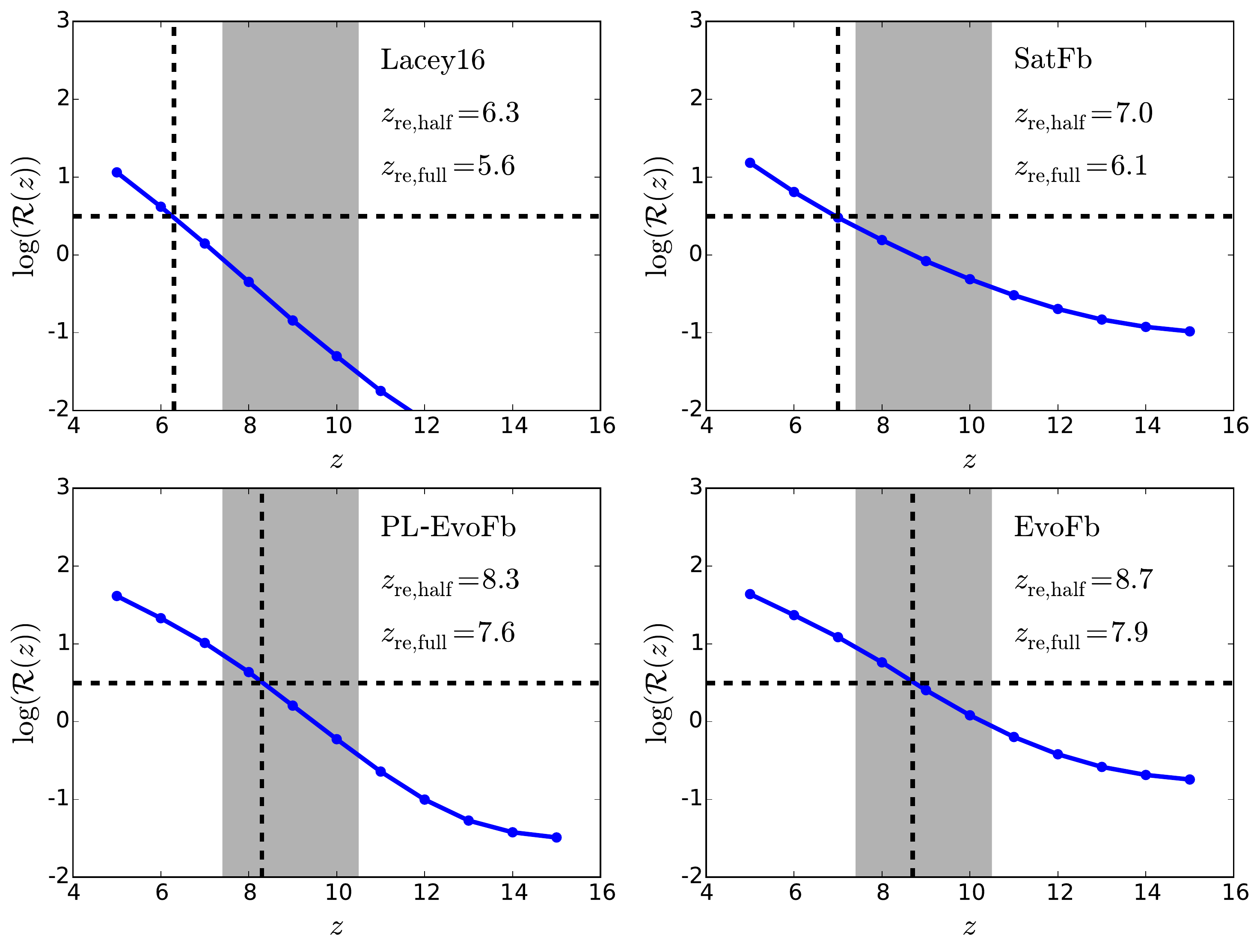}
\caption{${\cal R}(z)$, which is the ratio of the total number of
  ionizing photons produced up to redshift $z$ to the total number of
  hydrogen nuclei, for different SN feedback models. Each panel shows
  a different model, as labelled. The blue line shows the predicted
  ${\cal R}(z)$, while the horizontal dashed line shows the threshold
  ${\cal R}(z_{\rm re,half})=5$ for 50\% reionization, and the vertical dashed
  line the corresponding redshift $z_{\rm re,half}$. The grey shaded
  region shows the observational constraint on $z_{\rm re,half}$ from
  \Planck, namely $z_{\rm re}=8.8_{-1.4}^{+1.7}$ ($68\%$ confidence
  region, \citealp{plank2015_zre}). The predicted value of the
  redshift $z_{\rm re,full}$ for 100\% reionization is also given in
  each panel. }
\label{fig:z_re}
\end{figure*}

Fig.~\ref{fig:z_re} shows the predicted ${\cal R}(z)$ (defined in
Eq~\ref{eq:reion1}) for different SN feedback models. In each panel, 
the horizontal black dashed line indicates the criterion
for 50\% reionization, i.e.\ ${\cal R}(z_{\rm re,half})=3.125$, the
vertical black dashed line indicates $z_{\rm re,half}$ of the
corresponding model, and the corresponding value of $z_{\rm re,half}$
is given in the panel. The gray shaded area in each of these panels
indicates the current observational constraint from \Planck, namely
$z_{\rm re}=8.8_{-1.4}^{+1.7}$ ($68\%$ confidence region,
\citealp{plank2015_zre}). The redshift $z_{\rm re,full}$ for full
reionization (given by ${\cal R}(z_{\rm re,full})=6.25$) for each model
is also given in the corresponding panel. The results for the Lacey16
model are shown in the upper left panel. With the above mentioned criterion,
this model predicts $z_{\rm re,half}=6.3$, which is too low compared
to the observational estimate. This indicates that in the Lacey16
model, star formation at high redshift, $z\gsim 8$, is suppressed too
much. There are two possible reasons for this oversuppression: one is
the SN feedback at high redshift is too strong, and the other is that
the SN feedback in low-mass galaxies is too strong (since the typical
galaxy mass is lower at higher redshift).

\begin{figure*} 
\centering
\includegraphics[width=1.0\textwidth]{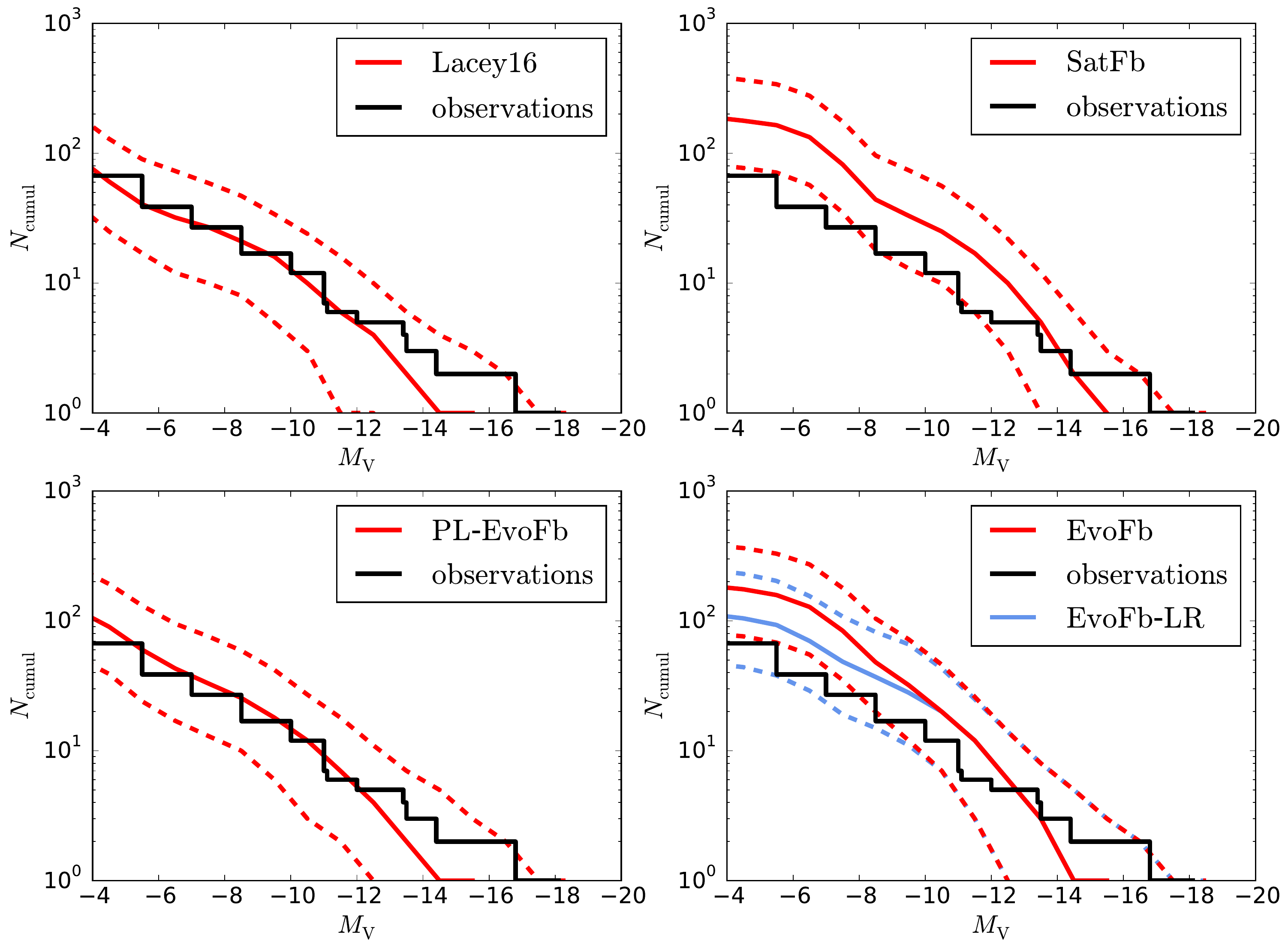}
\caption{Cumulative luminosity function of satellite
    galaxies in Milky Way-like host halos at $z=0$.  The solid black
    line in each panel is the observed Milky Way satellite cumulative
    luminosity function. For $M_{\rm V}<-11$, this shows the direct
    observational results from \citet{mcconnachie2012}, while for
    $M_{\rm V}\ge -11$, it shows the results from \citet{koposov2008},
    who applies some corrections for incompleteness in the
    observations. The other lines in each panel show the model
    predictions, with the solid red line showing the median for a
    sample of MW-like halos, and the dashed lines indicating the $5\--
    95\%$ range. The corresponding model names are given in the line
    key in each panel. }

\label{fig:mw_lf}
\end{figure*}

Fig.~\ref{fig:mw_lf} shows the cumulative luminosity function of
satellite galaxies in Milky Way-like host halos. In each panel, the
red solid and dashed lines show the simulation results for the
corresponding model. For each model, the simulations were run on 100
separate merger trees for each of 5 host halo masses, evenly spaced in
the logarithm of the mass in the range $5\times 10^{11}\--2\times
10^{12}\Msol$. This simulated sample of MW-like halos contains $500$
halos in total, and the red solid line shows the median satellite LF
for this sample, while the red dashed lines indicate the $5\-- 95\%$
range. The black solid line in each panel shows the observed Milky Way
satellite luminosity function. For $M_{\rm V}<-11$, we plot the direct
observational measurement from \citet{mcconnachie2012}. For these
brighter magnitudes, current surveys for MW satellites are thought to be
complete over the whole sky. For $M_{\rm V}\ge -11$ we plot the
observational estimate from \citet{koposov2008} based on SDSS, which
includes corrections for incompleteness due to both partial sky
coverage and in detecting satellites in imaging data. The predictions
for the Lacey16 model are shown in the upper left panel, and are in
very good agreement with the observations.

\begin{figure*} 
\centering
\includegraphics[width=1.0\textwidth]{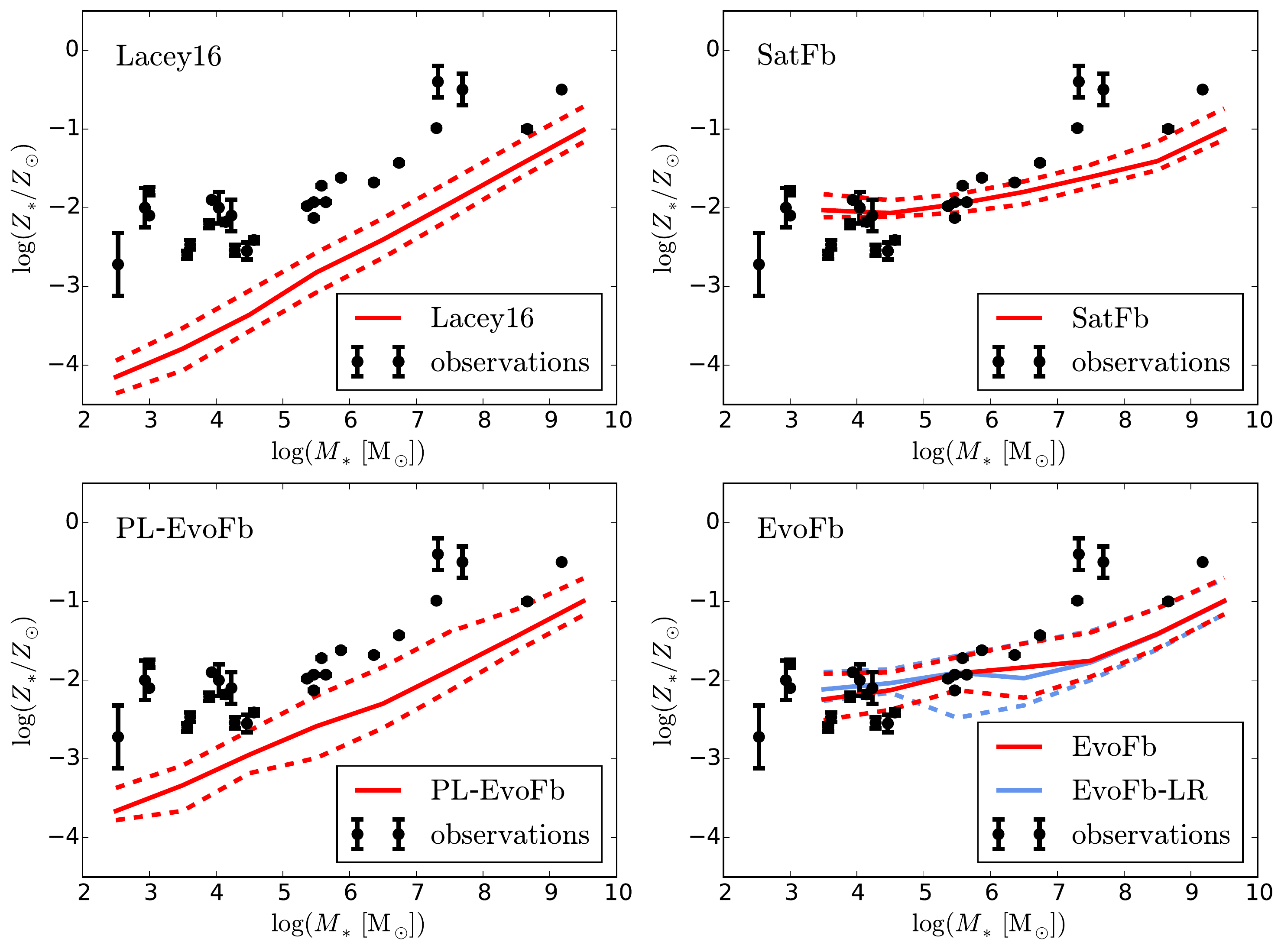}
\caption{The stellar metallicity ($Z_*$) vs stellar mass ($M_*$)
  relation for satellite galaxies in Milky Way-like host halos at
  $z=0$. The simulated sample for each moel is the same as in
  Fig.~\ref{fig:mw_lf}. In each panel, the solid line shows the
  median of the simulated sample, while the dashed lines indicate
  the $5\-- 95\%$ range, and the corresponding model name is given in the line key. The black filled circles show
  the observational results compiled by \citet{mcconnachie2012}. The
  observed $[{\rm Fe/H}]$ values in \citet{mcconnachie2012} are
  converted into the total stellar metallicities, $Z_*/\Zsol$, in Solar
  units by assuming the chemical patterns of the observed satellites
  are Solar. The total metallicities, $Z_*$, predicted by the model,
  which are absolute values, are converted into Solar units assuming
  $\Zsol=0.0142$ \citep{Asplund09}.}

\label{fig:mw_metal}
\end{figure*}

Fig.~\ref{fig:mw_metal} shows the $Z_*\-- M_*$ correlation for
satellite galaxies in Milky Way-like host halos. The sample is the
same as that for Fig.~\ref{fig:mw_lf}. In each panel, the red solid
line shows the median of the sample, while the red dashed lines
indicate the $5\-- 95\%$ range. The black filled circles in each
panel show observational data. We have converted the observed
$[{\rm Fe/H}]$ values 
into the total stellar metallicities, $Z_*/\Zsol$, by assuming that the
chemical abundance patterns in the observed satellites are the same as
in the Sun. This assumption may lead to an underestimation of the
metallicities of low mass satellites, which may not have had enough
enrichment by Type Ia supernova to reach the Solar pattern. For these
satellites, the observed $Z_*$ values shown in the figure are
therefore effectively lower limits. The results of the Lacey16 model
are again shown in the upper left panel. The $Z_*\-- M_*$ relation
predicted by this model is about an order of magnitude below the
observations.
Because the discrepancy in metallicity is about one order of
magnitude, it cannot be caused by inaccuracies in the theoretical
stellar yields of metals in this model or by the
variation of these yields with stellar metallicity. These yields
are obtained by integrating the yields predicted by stellar evolution
models over the IMFs assumed for stars formed either quiescently or in
starbursts. Assuming that the true metal yields are similar to what is
assumed in the model, then for a given stellar mass, the total metals
produced are fixed, so the low metallicities seen in the Lacey16 model
imply that the loss of metals from satellite galaxies is excessive.
Since the metal loss is caused by the outflows induced by SN feedback,
this indicates that the SN feedback in these small galaxies is too
strong.

In summary, the Lacey16 model motivates two types of modification to
the SN feedback. One is suppressing SN feedback in small galaxies,
which is the saturated feedback model. The other one is suppressing SN
feedback at high redshift, i.e.\ $z\ge 8$, but keeping strong feedback
at $z<4$ in order to reproduce the $z=0$ field LFs. This corresponds
to the evolving feedback model. Below, these two kinds of modification
will be tested one at a time.

\subsection{Saturated feedback model (SatFb)} 
The dashed green lines in Fig.~\ref{fig:field_lf} show the ${\rm
  b}_{\rm J}$-band and ${\rm K}$-band LFs for the SatFb model. These
predictions are still roughly consistent with the observations, but a
small excess of galaxies begins to appear at the very faint end of the
${\rm b}_{\rm J}$-band LF ($M_{{\rm b}_{\rm J}}>-17$). Reducing the SN
feedback strength further in this model would exacerbate this
discrepancy.

The upper right panel of Fig.~\ref{fig:z_re} shows ${\cal R}(z)$ for
our SatFb model; the predicted $z_{\rm re,half}$ is $7.0$, outside the
1-$\sigma$ region allowed by the \Planck observations. Thus, SN
feedback in the SatFb model is much too strong to allow production of
enough ionizing photons to reionize the Universe early enough.  The
upper right panel of Fig.~\ref{fig:mw_lf} shows the satellite LF of
Milky Way-like galaxies in the SatFb model. The relatively weak SN
feedback in this model leads to an overprediction of faint ($M_{\rm
  V}\ge -8$) satellites. Bearing in mind the significant uncertainties
in the numbers of faint satellites, this model prediction may be
deemed to be roughly acceptable.  Furthermore, these very faint Milky
Way satellites are very small, and so their abundance could be further
supressed by adjusting the strength of photoionization
feedback. However, this would not help reduce the excess at the faint
end of the field LFs, because these galaxies are larger and thus not
strongly affected by photoionization feedback.  The upper right panel
of Fig.~\ref{fig:mw_metal} shows the satellite $Z_*\-- M_*$
correlation for Milky Way-like hosts. This model prediction agrees
with observations only roughly.  The correlation is shallow because
most of these satellites have $\Vc< \Vthre = 62\kms$, and thus similar
values of $\beta$.

If the SN feedback strength in the SatFb model were further reduced,
the excess in the satellite LF would shift to brighter luminosities,
$M_{\rm V}<-8$, where there are fewer uncertainties in the data and
where photoionization feedback is ineffective. The stellar metallicity
of satellites of a given stellar mass would become even higher,
spoiling the already marginal agreement with observations. Together,
these results from the Milky Way satellites suggest that the strength
of SN feedback in model SatFb is a lower limit to the acceptable
value.

This SatFb model therefore does not provide a solution to the problems
identified in the Lacey16 model. Further adjustments within the
framework of the saturated feedback model would involve changing the
saturated power-law slope $\gammapSN$ and/or the threshold velocity,
$\Vthre$. In the present SatFb model, as mentioned above, $\gammapSN$
is already at its lower limit, namely $0$, and introducing a positive
$\gammapSN$ only leads to a stronger SN feedback in small galaxies
than in the current SatFb model, and this would not predict a high
enough $z_{\rm re,half}$. Reducing $\Vthre$ would also lead to a
stronger SN feedback in small galaxies than in the current SatFb
model, so would not improve the prediction for $z_{\rm re,half}$
either, while enhancing $\Vthre$ would lead to a saturation of the SN
feedback in even larger galaxies and a stronger saturation in small
galaxies than in the current SatFb model. Since the feedback strength
in the SatFb model is already as low as allowed by observations of the
field LFs, the MW sat LF and the MW sat $Z_*\-- M_*$ relation, this
adjustment would only worsen these discrepancies. Thus, the saturated
feedback model is disfavoured by this combination of observational
constraints.

\subsection{Evolving feedback model}
\subsubsection{PL-EvoFb model}
The magenta lines in Fig.~\ref{fig:field_lf} show the ${\rm b}_{\rm
  J}$-band and ${\rm K}$-band field LFs for the PL-EvoFb model. The
results are very close to those in the Lacey16 model, and the observed
faint ends are well reproduced. This is because in the PL-EvoFb model,
the SN feedback at $z\le 4$ is the same as in the Lacey16 model. The
lower left panel in Fig.~\ref{fig:z_re} shows ${\cal R}(z)$ for this
model; the corresponding $z_{\rm re,half}$ is $8.3$, which is in
agreement with observations. This shows that the evolving feedback
model is more successful at generating early reionization than the
saturated feedback model.

The lower left panel of Fig.~\ref{fig:mw_lf} shows the
  satellite luminosity function of Milky Way-like host halos in the
  PL-EvoFb model, which is in very good agreement with the
  observations. The lower left panel of Fig.~\ref{fig:mw_metal} shows
  the $Z_*\-- M_*$ relation for satellite galaxies in Milky
  Way-like host halos in this model. This model predicts stellar
  metallicities of satellites with $M_*\leq 10^6\,{\rm M}_{\odot}$
  several times to one order of magnitude lower than observations,
  with the discrepancy increasing with decreasing stellar
  mass. Although weakening the SN feedback at high redshifts does
  improve the result compared to the Lacey16 model, it is still
  inconsistent with observations. Thus this model is disfavoured by
  observations of MW satellite metallicities. The discrepancy again
  suggests that the SN feedback in small galaxies is too strong, but
  since at the same time this model successfully reproduces the faint
  ends of the field LFs, it suggests that this problem of too strong
  feedback is restricted to very small galaxies. This then motivates
  our next model, in which we preferentially suppress the SN feedback
  strength in very small galaxies, while retaining the same evolution
  of feedback strength with redshift as in the PL-EvoFb model.

\subsubsection{EvoFb model}
The field LFs predicted by the EvoFb model are almost
  identical to those given by the PL-EvoFb model, so this model
  likewise successfully reproduces the faint ends of the field
  LFs. The reason for the similarity between the field LFs predicted
  by the two models is that the saturation introduced in the EvoFb
  model is only effective for $\Vc \leq 50\kms$, and would not
  significantly affect the galaxies in the observed faint ends of the
  field LFs, which typically have higher $\Vc$.

  The lower right panel in Fig.~\ref{fig:z_re} shows ${\cal R}(z)$ for
  the EvoFb model; the corresponding $z_{\rm re,half}$ is $8.7$,
  which is in agreement with the observations. Compared to the result of
  the PL-EvoFb model, $z_{\rm re,half}$ only increases slightly, so
  the saturation in the feedback has only a small effect, and the main
  factor leading to the agreement with observations is still the
  redshift evolving behavior of the SN feedback strength.

The lower rigth panel of Fig.~\ref{fig:mw_lf} shows
  the satellite luminosity function of Milky Way-like host halos in
  the EvoFb model. The model predictions are roughly consistent with
  the observations, although the very faint end ($M_{\rm V}\ge -8$) of
  the MW sat LF is somewhat too high. However, as mentioned in
  connection with the SatFB model, the observations of this very faint end have
  significant uncertainties, so this model is still acceptable. The
  lower right panel of Fig.~\ref{fig:mw_metal} shows the $Z_*\-- M_*$
  relation for the satellite galaxies in Milky Way-like host halos in
  the EvoFb model. The model predictions are now roughly consistent
  with the observations. This improvement is achieved by adopting both
  an evolving SN feedback strength and a saturation of the feedback in
  galaxies with $\Vc\le 50\kms$.

Because the predictions for Milky Way satellites are sensitive to the
photoionization feedback, it is possible to further improve the
agreement with observations for these galaxies by adjusting the
photoionization feedback. One possible adjustment is to adopt the
so-called local reionization model (see \citet{font2011} and
references therein), in which higher density regions reionize earlier,
so that $z_{\rm re,full}$ for the Local Group region is earlier than
the global average $z_{\rm re,full}$ constrained by the \Planck data.
Earlier reionization means earlier photoionization feedback, so that
for the Milky Way satellites one has $\zcrit>z_{\rm
  re,full}$. \citet{font2011} adopted a detailed model to study this
local reionization effect, and suggested that using $\zcrit=10$ gives
a good approximation to the results of the more detailed model. Here
we also adopt $\zcrit=10$, and we label the model with evolving SN
feedback and $\zcrit=10$ as EvoFb-LR.

We tested that the predictions for global properties like $z_{\rm
  re,full}$, $z_{\rm re,half}$ and the field LFs are not very
sensitive to the value of $\zcrit$. It is therefore justified to
ignore the variation of $\zcrit$ with local density when calculating
these global properties, and adopt a single $\zcrit=z_{\rm re,full}$
when predicting these.
This also means that introducing such a local reionization model does
not allow one to bring the standard Lacey16 or SatFb models into
agreement with all of our observational constraints, since some of the
discrepancies described above involve these global properties.

The satellite luminosity function of the Milky
  Way-like host halos in the EvoFb-LR model is also shown in the lower
  right panel of Fig.~\ref{fig:mw_lf}. The model predictions agree
  with observations better than the EvoFb model, because the abundance
  of the very faint satellites is reduced by the enhanced
  photoionization feedback. The $Z_*\-- M_*$ relation for satellite
  galaxies in Milky Way-like host halos for the EvoFb-LR model is very
  similar to that of the EvoFb model, shown in
  Fig.~\ref{fig:mw_metal}.


\section{discussion}
\label{sec:discussion}

\subsection{Why should the SN feedback strength evolve with redshift?}
The physical idea behind formulating the mass loading factor, $\beta$,
of SN-driven outflows (Eq \ref{eq:beta_def}) as a function of $\Vc$
is that the strength of the SN feedback driven outflows (for a given
star formation rate, $\psi$) depends on the gravitational potential
well, and $\Vc$ is a proxy for the depth of the gravitational
potential well. However, in reality the strength of outflows does not
only depend on the gravitational potential well, but may also depend
on the galaxy gas density, gas metallicity and molecular gas
fraction. This is because the gas density and metallicity determine
the local gas cooling rate in the ISM, which determines the fraction
of the injected SN energy that can finally be used to launch outflows,
while the dense molecular gas in galaxies may not be affected by the
SN explosions, and thus may not be ejected as outflows. These
additional factors may evolve with redshift, and $\Vc$ may not be a
good proxy for them, so if the outflow mass loading factor, $\beta$, is
still formulated as a function of $\Vc$ only, a single function may
not be valid for all redshifts and some redshift evolution of $\beta$
may need to be introduced.

The detailed dependence of $\beta$ on the galaxy gas density, gas
metallicity and molecular gas fraction can only be derived by using a
model which considers the details of the ISM. The model of
\citet{lagos_feedback} is an effort towards this direction, and the
dependence of $\beta$ on $\Vc$ predicted by that model is shown in
Fig.~15 of that paper. But since the model in \citet{lagos_feedback}
only considers ejecting gas out of galaxies, but does not predict what
fraction of this escapes from the halo, the model is incomplete. We
therefore only use very general and rough features of the dependence
of $\beta$ on $\Vc$ and $z$ predicted by \citet{lagos_feedback} to
motivate our PL-EvoFb and EvoFb models, which assume
  a redshift-dependent $\beta$.

\citet{lagos_feedback} suggest that the mass loading, $\beta$, is weaker
in starbursts than for quiescent star formation in galaxy disks,
because starbursts have higher gas density and molecular gas
fraction. While this feature is not included in our model, as it may 
be too complex for a phenomenological SN
feedback models, it has the potential to enhance the
reionization redshift and the stellar metallicities of galaxies, so it
might be worth investigating it in future work.

\subsection{What kind of galaxies reionized the Universe?}

\begin{figure*} 
\centering
\includegraphics[width=1.0\textwidth]{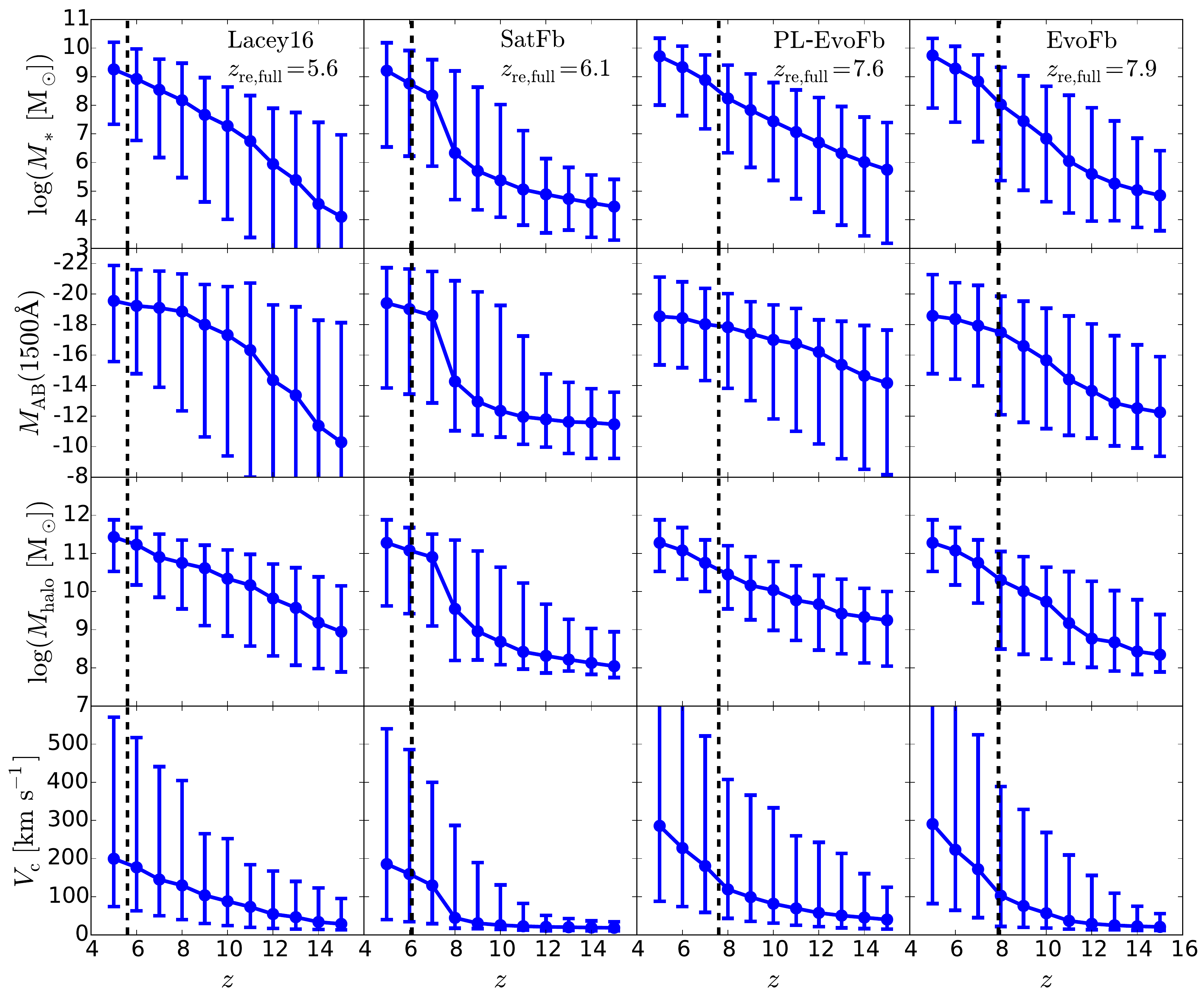}
\caption{Simple statistics of the galaxies producing ionizing
  photons. Each column corresponds to a different SN feedback model,
  with the corresponding model name given in the top of each
  panel in the first row, along with the value of $z_{\rm re,full}$,
  the redshift at which the Universe is fully ionized. The vertical
  dashed lines in each panel also indicate $z_{\rm re,full}$. The
  first row shows the statistics of the stellar mass, $M_*$, of the
  galaxies producing ionizing photons, the second row shows the
  statistics of the dust-extincted rest frame UV magnitude, $M_{\rm
    AB}(1500{\rm \AA})$, of these galaxies, while the third row shows
  the statistics of the halo masses, $M_{\rm halo}$ and the fourth row
  shows the statistics of the galaxy circular velocity, $\Vc$. For
  each quantity shown in these rows, the dots indicate
  the medians of the corresponding quantity, and the errorbars
  the $5\-- 95\%$ range, with both the medians and the
  $5\-- 95\%$ ranges being determined by their contributions to the
  ionizing photon emissivity at that redshift.}
\label{fig:re_info}
\end{figure*}

\begin{figure*} 
\centering
\includegraphics[width=1.0\textwidth]{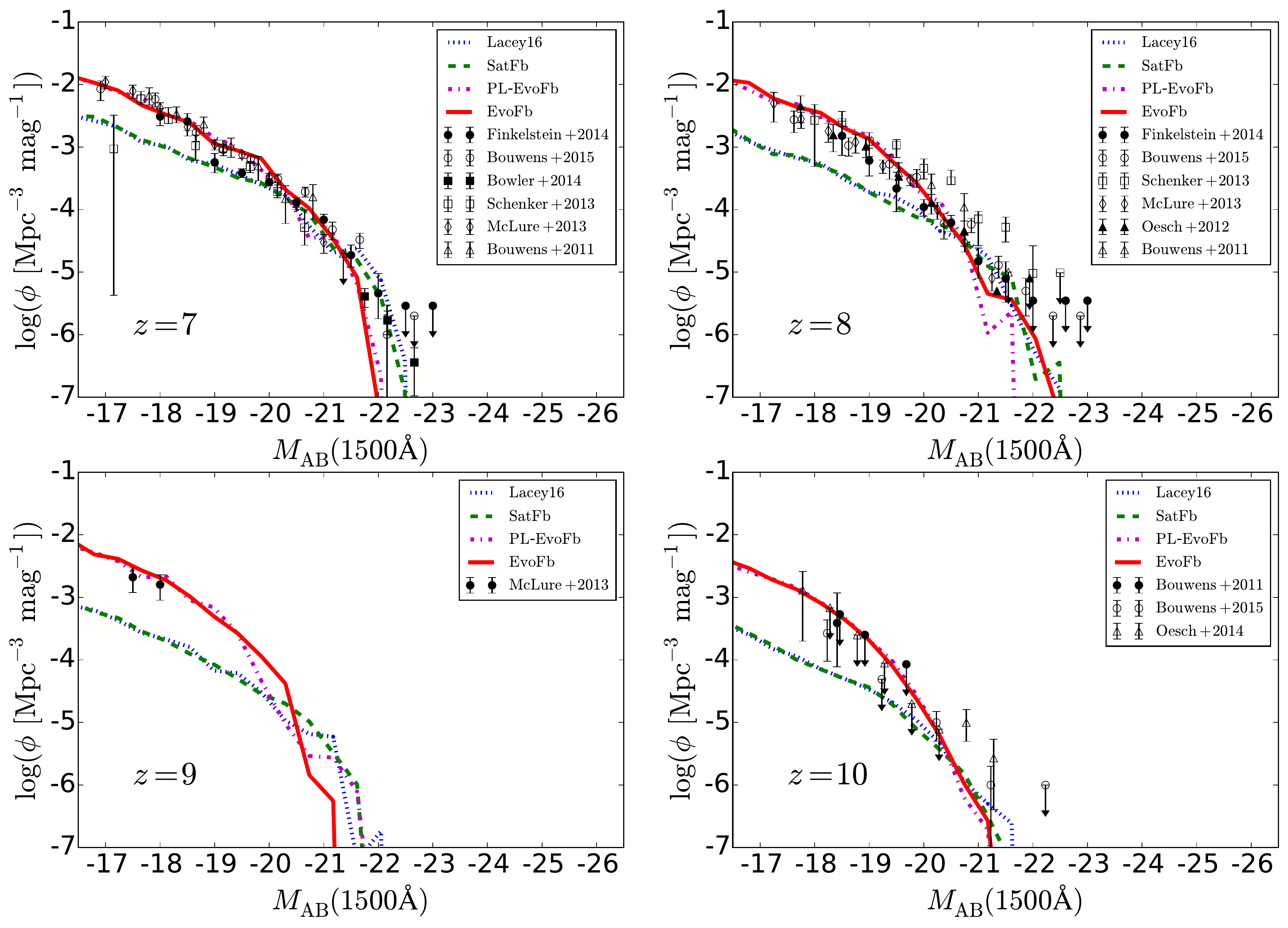}
\caption{The rest frame far-UV luminosity functions at $z=7,\ 8,\ 9,\
  10$ for the 4 different SN feedback models. In each
  panel, the blue dotted line shows the prediction for the Lacey16
  model, the dashed green line that for the SatFb model,
  the magenta dotted dashed line that for the PL-EvoFb
    model and the red solid line that for the EvoFb model, while
  symbols with errorbars indicate observational measurements
  \citep{uv_lf_bouwens2011, uv_lf_bouwens2011n, uv_lf_oesch2012,
    uv_lf_schenker, uv_lf_mclure, uv_lf_finkelstein, uv_lf_bowler,
    uv_lf_oesch2014, uv_lf_bouwens2015}. The
    dust extinction is calculated self-consistently based on galaxy gas
    content, size and metallicity (see \citet{Lacey16_model} for more
    details).
}
\label{fig:uv_lf}
\end{figure*}

\begin{figure*} 
\centering
\includegraphics[width=0.7\textwidth]{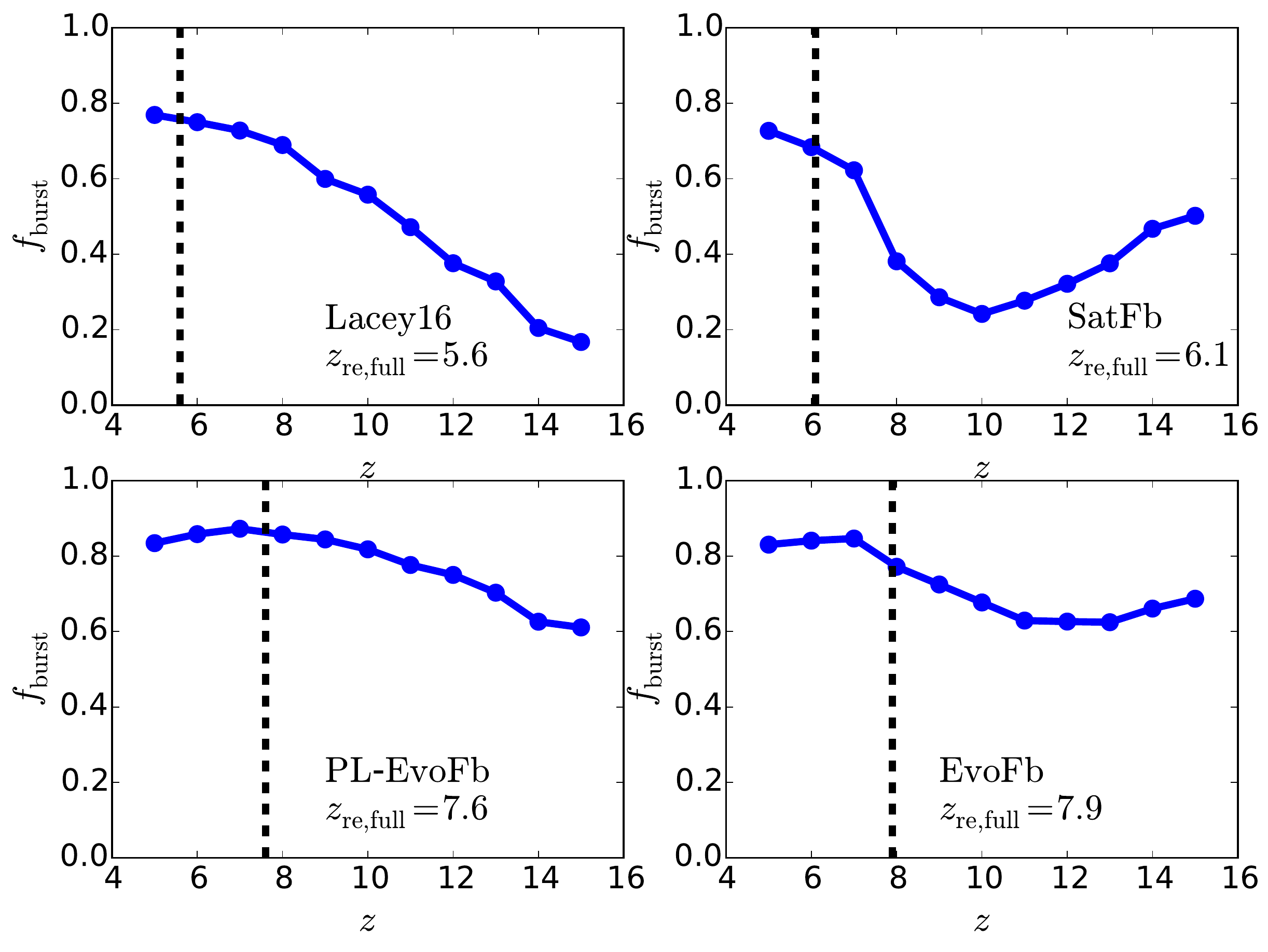}
\caption{The fraction of the ionizing photon emissivity contributed by
  starbursts at a given redshift. Different panels are for different
  SN feedback models, as labelled, and the vertical dashed lines
  indicate $z_{\rm re,full}$.}
\label{fig:re_f_burst}
\end{figure*}

Fig.~\ref{fig:re_info} shows some simple statistics of the galaxies
producing the ionizing photons. The first row shows the statistics of
the stellar mass, $M_*$, of the galaxies producing ionizing photons,
the second row shows the statistics of the dust-extincted rest-frame
far-UV absolute magnitude, $M_{\rm AB}(1500{\rm \AA})$, of these
galaxies, while the third row shows the statistics of the halo masses,
$M_{\rm halo}$, and the fourth row the statistics of the galaxy
circular velocity, $\Vc$. For each quantity, the dots in each panel
indicate the medians of the corresponding quantity, and the error bars
indicate the $5\-- 95\%$ range, with the medians and percentiles
determined not by the number of galaxies but by their contributions to
the ionizing emissivity at that redshift. The median means that
galaxies below it contribute $50\%$ of the ionizing photons at a given
redshift, while the $5\-- 95\%$ range indicates that the galaxies
within it contribute $90\%$ of the ionizing photons at a given redshift. Each
column corresponds to a different SN feedback model. The vertical
dashed lines in each panel indicate $z_{\rm re,full}$, the redshift at
which the Universe is fully ionized, for that model, with the
numerical values of $z_{\rm re,full}$ given in the panels in the
first row.

From Fig.~\ref{fig:re_info} it is clear that the median of $M_*$ at
$z\sim z_{\rm re,full}$ for each SN feedback model is around
$10^8\--10^9\Msol$, the median of $M_{\rm AB}(1500{\rm \AA})$ is
around $-17\-- -19$, and the median of $\Vc$ is around $100\--
200\kms$. These values indicate that the corresponding galaxies are
progenitors of large massive galaxies at $z=0$. This means in these
models, the progenitors of large galaxies make significant
contributions to the cosmic reionization. It is also true that the
progenitors of large galaxies have already made contributions to
the ionizing photons when the Universe was half ionized, i.e.\ by
$z=z_{\rm re,half}$. This means that a preferential suppression of the
SN feedback in very small galaxies is not very effective in boosting
$z_{\rm re,half}$, and to predict a high enough $z_{\rm re,half}$ by
these means usually requires heavy suppression of the SN feedback in
very small galaxies, which spoils the agreement with observations of
faint galaxies at $z=0$. This is the reason for the failure of the
SatFb model to satisfy all the observational constraints considered in
this work.

Fig.~\ref{fig:re_info} also shows that the median of $M_{\rm halo}$ in
each SN feedback model is roughly in the range $10^{10}\--
10^{11}\Msol$ at $z\sim z_{\rm re,full}$, which means there are
significant contributions to the ionizing photons from large atomic
hydrogen cooling halos. This is consistent with the results from
\citet{atomic_halo_reion}, who show that it is difficult to obtain
reionization at $z\sim 8$ mainly from star formation in small atomic
cooling halos with $M_{\rm halo}\sim 10^8\Msol$.

We also calculated the rest-frame far-UV luminosity functions at
$z=7,\ 8,\ 9,\ 10$ for our 4 different SN feedback models. These
predictions are shown in Fig.~\ref{fig:uv_lf}, and compared with
recent observational data. The best fit (EvoFb) model is seen to agree
quite well with the observations over the whole range $z=7 \-- 10$. 
The PL-EvoFb model, which adopts similar redshift evolving SN feedback, 
also reaches similar level of agreement with observations.
On the other hand, the other 2 models, which generally have stronger SN
feedback at high redshift than the EvoFb model, predict too few low UV
luminosity galaxies at $z = 7 \-- 10$.  Note that the current
observational limit is $M_{\rm AB}(1500{\rm \AA})\sim -17 \-- -18$ at
these redshifts, which is close to the median of $M_{\rm AB}(1500{\rm
  \AA})$ at reionization for the EvoFb model shown in
Fig.~\ref{fig:re_info} (for this model, at $z=8$, the median is
$M_{\rm AB}(1500{\rm \AA})=-17.5$, and the $5\-- 95\%$ range is
$M_{\rm AB}(1500{\rm \AA})=-12.1$ to $M_{\rm AB}(1500{\rm
  \AA})=-19.8$.). Thus the best fit model suggests that the currently
observed high redshift galaxy population should contribute about half
of the ionizing photons that reionized the Universe. 
This is consistent with \cite{reion_threo2}, which 
suggests that the sources of reionization can not be too heavily dominated by very faint galaxies.

We also checked that the rest-frame far-UV luminosity
  functions predicted by all 4 models become very similar at $z\leq
  6$, and thus the modifications to the SN feedback do not spoil the
  good agreement of these luminosity functions with observations at
  $3\leq z \leq 6$ found in the original Lacey16 model.

  Fig.~\ref{fig:re_f_burst} shows the fraction of the ionizing photons
  that are contributed by starbursts at a given redshift (as compared
  to stars formed quiescently in galaxy disks). Different panels are
  for different SN feedback models, and the vertical dashed lines
  indicate $z_{\rm re,full}$ for the corresponding models. It is clear
  that at $z\sim z_{\rm re,full}$, the starburst fractions are high,
  with $f_{\rm burst}\approx 0.8$ in all four models. This indicates
  that starbursts are a major source of the ionizing photons for
  cosmic reionization.

\subsection{The descendants of the galaxies that ionized Universe}

\begin{figure} 
\centering
\includegraphics[width=7.0 cm]{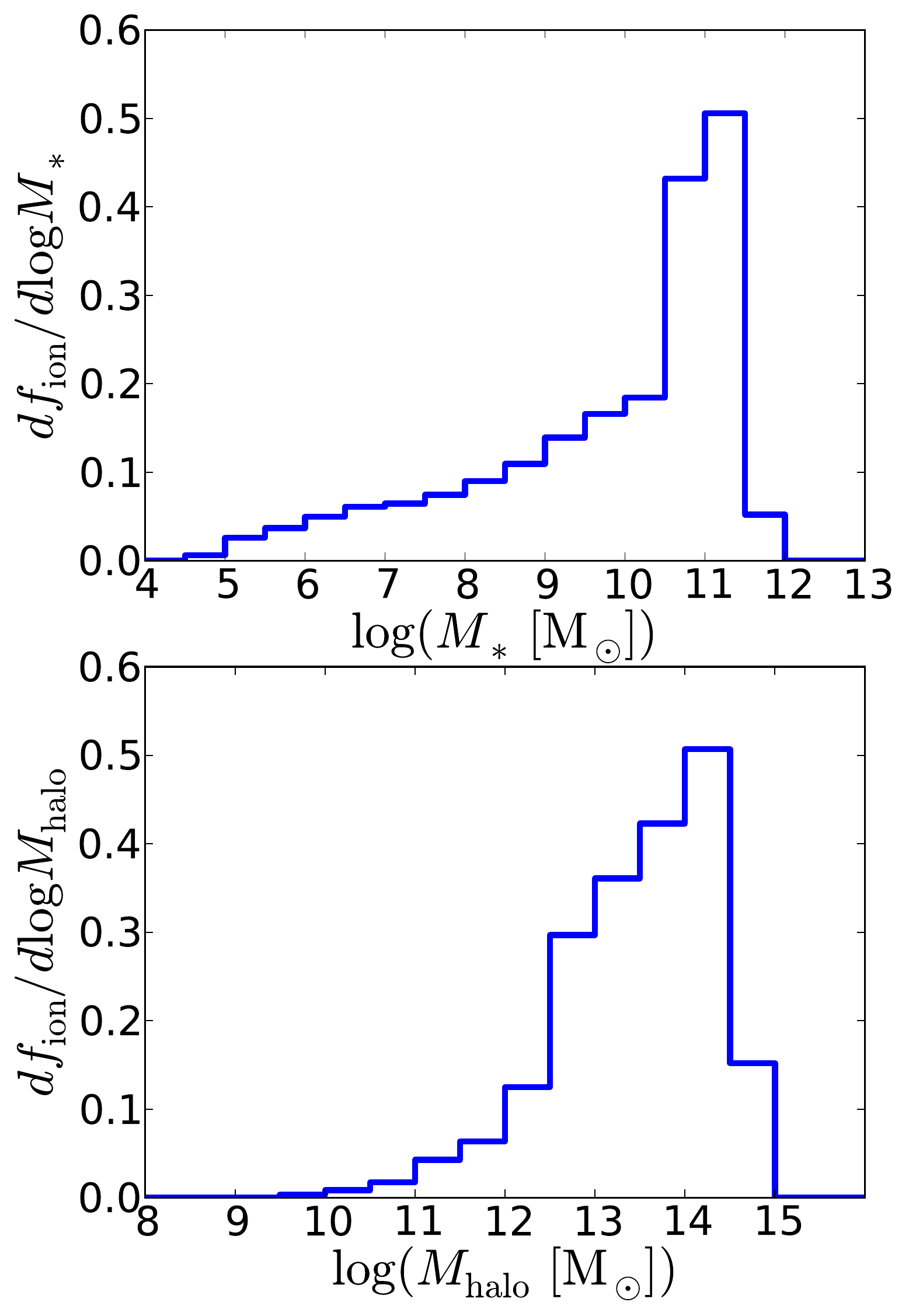}
\caption{Probability distributions of masses of $z=0$ descendants of
  objects which emit ionizing photons at $z \geq z_{\rm re,full}$,
  weighted by number of ionizing photons produced.  {\bf Upper panel:}
  Probability distribution of stellar mass of descendant at
  $z=0$. {\bf Lower panel:} Probability distribution of halo mass of
  descendant at $z=0$.}
\label{fig:f_ion}
\end{figure}

For the best fit model, i.e.\ the EvoFb model, we also identified the
$z=0$ descendants of the galaxies which ionized the Universe. To do
this, we ran a simulation with fixed dark matter halo mass resolution
$M_{\rm res}=7.1\times 10^7\Msol$ from $z=20$ to $z=0$. This $M_{\rm
  res}$ is low enough to ensure that we resolve all the atomic cooling
halos up to $z=11$. According to Fig.~\ref{fig:z_re}, most of the
ionizing photons that reionized the Universe are produced near $z_{\rm
  re,full}$, and for the EvoFb model, $z_{\rm re,full}=7.9$. Thus,
resolving all the atomic cooling halos up to $z=11$ ensures that all
galaxies which are major sources of the ionizing photons and their
star formation histories are well resolved.

In Fig.~\ref{fig:f_ion} we show the mass distributions of the $z=0$
descendants of the objects which produced the photons which reionized
the Universe, weighted by the number of ionizing photons produced. The
top panel shows the stellar mass of the descendant galaxy, while the
bottom panel shows the mass of the descendant dark matter halo. To
calculate these, we effectively identify each ionizing photon emitted
at $z \geq z_{\rm re,full}$, then identify the $z=0$ descendant
(galaxy or halo) of the galaxy which emitted it, then construct the
probability distribution of descendant mass, giving equal weight to
each ionizing photon. The upper panel of Fig.~\ref{fig:f_ion} shows
that over $50\%$ of the ionizing photons are from the progenitors of
large galaxies with $M_* > 3 \times 10^{10} \Msol$, or equivalently,
the major ionizing sources have $z=0$ large galaxies as their
descendants. The lower panel of Fig.~\ref{fig:f_ion} shows that $50\%$
of the ionizing photons are from the progenitors of high mass dark
matter halos at $z=0$ with $M_{\rm halo} > 3.7 \times 10^{13} \Msol$,
which means that the reionization is driven mainly by sources at very
rare density peaks.  These results are consistent with the indications
given by Fig.~\ref{fig:re_info}.

\begin{figure} 
\centering
\includegraphics[width=7.0 cm]{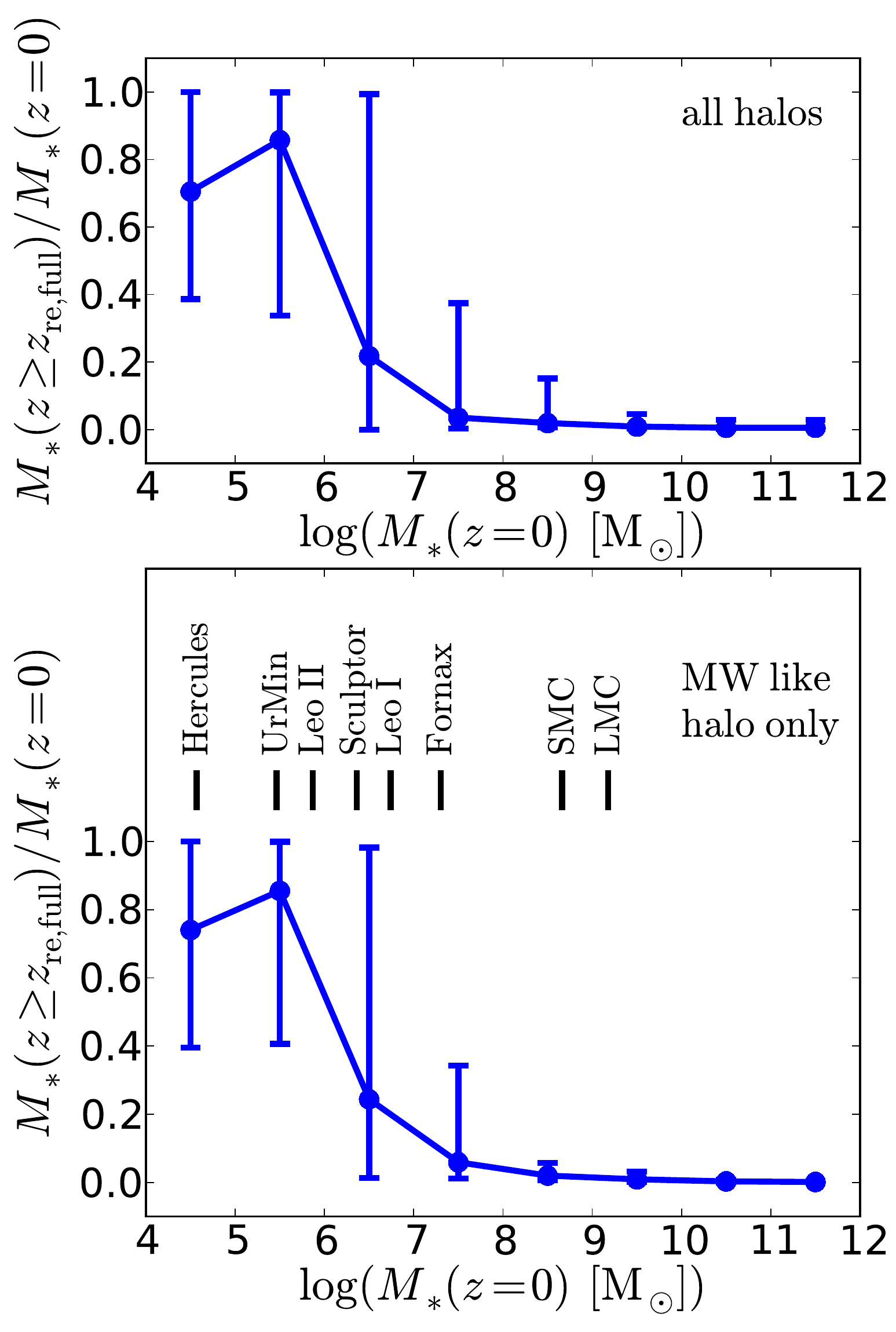}
\caption{Fraction of stellar mass in galaxies at $z=0$ which was
  formed before reionization (i.e.\ at $z\ge z_{\rm re,full}$). In
  both panels, each filled circle shows the median of the ratio in the
  corresponding $z=0$ stellar mass bin, while the corresponding error
  bar indicates the $5\-- 95\%$ range of this ratio. {\bf Upper
    panel:} All galaxies. {\bf Lower panel:} Galaxies in Milky
  Way-like halos only (defined as halos with $z=0$ halo mass in the
  range $5\times 10^{11}\Msol \le M_{\rm halo} \le 2\times
  10^{12}\Msol$). The short vertical solid black lines indicate the
  observed stellar masses of several Milky Way satellites, namely LMC,
  SMC, Fornax, Sculptor, Leo I, Leo II, Ursa Minor (UrMin) and
  Hercules, for reference (from \citealt{mcconnachie2012}).}
\label{fig:f_star}
\end{figure}

In Fig.~\ref{fig:f_star}, we show the fraction of stellar mass in
galaxies at $z=0$ that was formed before reionization, i.e. at
$z\ge z_{\rm re,full}$, for the best fit model (the EvoFb model). The
upper panel shows this for all galaxies, while the lower panel shows
this quantity only for galaxies in Milky Way-like halos, defined as
halos with $z=0$ halo mass in the range $5\times 10^{11}\Msol \le
M_{\rm halo} \le 2\times 10^{12}\Msol$.  The upper panel shows that
even though the progenitors of the $z=0$ large galaxies provided about
half of the ionizing photons, only a tiny fraction of their stars are
formed before reionization, and while the $z=0$ dwarf galaxies
($M_*(z=0)<10^6\Msol$) contributed only a small fraction of the
photons for reionization, their stellar populations typically are
dominated by the stars formed before reionization. This is consistent
with the hierarchical structure formation picture, because smaller
objects formed earlier, and also formation of galaxies in small halos
is suppressed after reionization by photoionization feedback. Also
note that the ratio of the mass of the stars formed at $z\ge z_{\rm
  re,full}$ to the $z=0$ stellar mass shows considerable scatter for
galaxies with $M_*(z=0)<10^7\Msol$, which means the star formation
histories of these small galaxies are very diverse.

The lower panel of Fig.~\ref{fig:f_star} shows galaxies in Milky
Way-like halos only, but the predicted fraction of stars formed before
reionization is in fact very similar to the average over all halos
shown in the upper panel. For reference, the short vertical solid
black lines indicate the observed stellar masses of several Milky Way
satellites (from \citealt{mcconnachie2012}), namely LMC, SMC, Fornax,
Sculptor, Leo I, Leo II, Ursa Minor (UrMin) and Hercules. As shown by
this panel, the best fit model implies that for the large satellites
like the LMC, SMC and Fornax, only tiny fractions of their stellar
mass, typically 5\% or less, were formed before reionization. However,
this fraction increases dramatically with decreasing satellite mass,
as does the scatter around the median. For the lowest mass satellites,
with stellar mass $M_*< 10^6\Msol$, including objects like Leo II,
Ursa Minor and Hercules, the median fraction increases to around 80\%,
meaning that most of the satellites in this mass range form the bulk
of their stars before reionization, with the 5--95\% range in this
fraction extending from 40\% to 100\%, indicating diverse star formation
histories for different satellites of the same mass. Satellites in the
intermediate mass range $10^6\Msol \le M_*< 10^7\Msol$, like Leo I and
Sculptor, have somewhat lower median fractions formed before
reionization, around 20--50\%, but with an even larger scatter around
this median, with the 5--95\% range extending nearly from 0\% to
100\%.

\subsection{Modelling uncertainties}
\label{sec:uncertainty}

An important assumption in our study is that $f_{\rm
    esc}$ is constant and, in our default model, equal to 0.2. This
  choice is justified in Section~\ref{sec:reionization_calculation};
  here we explore the effects of varying this parameter. We also
  explore the effect of including a contribution from AGN to the
  photoionizing budget, which in our standard model we assume to be
  negligible. 

\cite{AGN_reionization2} have recently revived the old
  idea that photons produced by AGN could be responsible for
  reionization. They take the observed AGN Lyman limit emissivity,
  $\epsilon_{912}$, at $z\leq 6$ and extrapolate it to $z\approx
  12$. Assuming that the AGN UV spectrum is a power law with index
  $-1.7$, they calculate the number of ionizing photons emitted by AGN
  per unit time per unit comoving volume, $\epsilon'_{\rm AGN}$. The
  redshift of reionization can then be obtained either by solving the
  equation for $Q_{\rm HII}$, or using the simpler method we
  introduced in
  \S\ref{sec:reionization_calculation}. \cite{AGN_reionization2}
  conclude that AGN alone could have been the dominant source of the
  photons respsonsible for reionization. 

\begin{figure*} 
\centering
\includegraphics[width=0.9\textwidth]{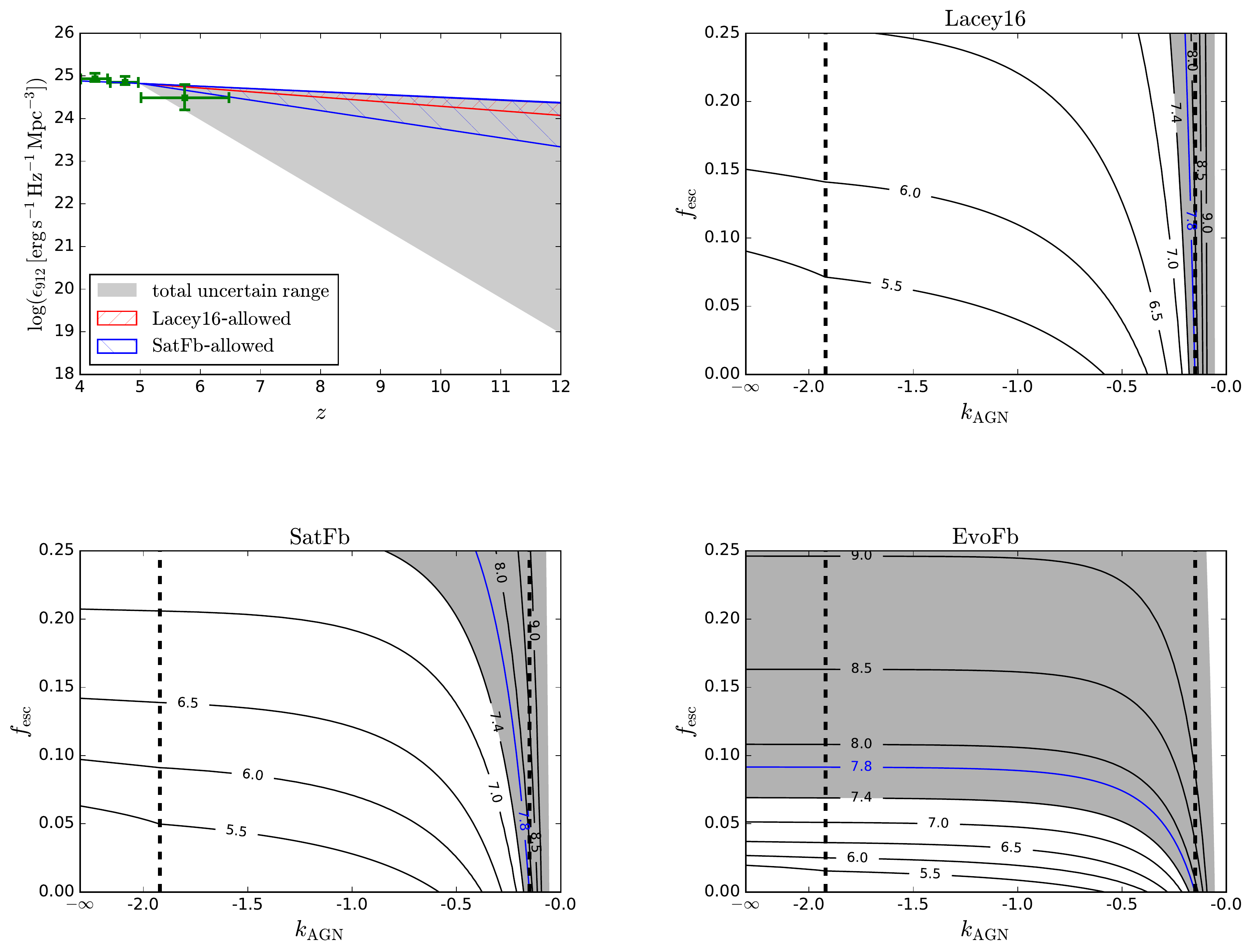}
\caption{{\bf Upper left panel:} extrapolations of the
    AGN emissivity at the Lyman limit, $\epsilon_{912}$, allowed by
    the errorbar of the measurement at $z\simeq 6$.  The three data
    points with errorbars are the observations taken from Fig.~1 of
    \citet{AGN_reionization2}, the grey shaded region shows the
    allowed extrapolations. The extrapolation adopted by
    \citet{AGN_reionization2} lies on the upper boundary of the
    region; the red and blue hatched regions encompass the
    extrapolations required to bring the Lacey16 and SatFb models
    respectively into agreement with the \Planck constraints, for
    $f_{\rm esc}=0.2$. Reducing $f_{\rm esc}$ shrinks these regions
    towards the upper boundary. {\bf Remaining three panels:}
    predicted $z_{\rm re,half}$ (contour lines) for different
    combinations of $f_{\rm esc}$ and $k_{\rm AGN}$, where $f_{\rm
      esc}$ is the escape fraction for stars and $k_{\rm AGN}$ is the
    slope of the AGN emissivity extrapolation shown in the upper left
    panel. The panels correspond to the Lacey16, SatFb and EvoFb
    models, as labelled.  The grey shaded area in each panel
    represents the region allowed at $1$-$\sigma$ by the \Planck data.
    The vertical dashed lines indicate the lower and upper limits of
    the extrapolation slope, i.e.\ $k_{\rm AGN}=-1.92$ and $k_{\rm
      AGN}=-0.15$. The line labelled $k_{\rm AGN}=-\infty$ corresponds
    to the case of no AGN contribution.}
\label{fig:uncertain1}
\end{figure*}

The estimate of $\epsilon_{912}$ at $z\approx 6$ has a
  large errorbar and so a major uncertainty in the model of
  \cite{AGN_reionization2} is their extrapolation to higher
  redshifts. They extrapolate using a complex functional form that,
  however, is close to an exponential, $\epsilon_{912}\propto \exp
  (\rm k_{\rm AGN}\, z)$, at $z\geq 5$, the regime relevant to
  hydrogen reionization. To assess the plausibility of the
  \cite{AGN_reionization2} model, we investigate other extrapolations
  of $\epsilon_{912}$, which are consistent with the measured value at
  $z=6$. We consider the same exponential form, constrained
  in all cases to lie within the errorbar of the measured value at
  $z=6$ and to give the same value at $z=5$ as the
  \cite{AGN_reionization2} model. (Unlike \citeauthor{AGN_reionization2},
  for simplicity, we extrapolate to $z=\infty$ rather than to $z\approx
  12$, as they do, but this overestimate of the AGN contribution 
  introduces only very small changes to the redshift of
  reionization.) These two requirements result in a family of
  extrapolated estimates, with $-1.92\leq k_{\rm AGN}\leq -0.15$,
  illustrated by the grey shaded region in the upper left panel of
  Fig.~\ref{fig:uncertain1}. The emissivity assumed by
  \cite{AGN_reionization2} lies at the upper boundary of this allowed
  region. Following \citeauthor{AGN_reionization2} we adopt an escape
  fraction of $100\%$ for AGN. There is considerable uncertainty on
  this parameter as well (see \cite{AGN_reionization2} for further
  discussion).

Once $\epsilon'_{\rm AGN}$ is known, the calculation in
  \S\ref{sec:reionization_calculation} can be extended to include
  AGN. Specifically, we have, 
\begin{eqnarray}
& \epsilon_{\rm tot}(z) & =  f_{\rm esc}\epsilon_{\rm star}(z)+\epsilon_{\rm AGN}(z) \\
& {\cal R}'(z) & =  \frac{\int_{z}^{\infty}\epsilon_{\rm tot}(z')\,dz'}{n_{\rm H}} \\
& {\cal R}'(z_{\rm re,full}) & =  1+N_{\rm rec},
 \label{eq:ext_zre_calculation}
\end{eqnarray}
where $\epsilon_{\rm star}$ is the emissivity of the stars, which is
given by \GALFORM, $f_{\rm esc}$ is the corresponding escape fraction,
$n_{\rm H}$ is the comoving number density of hydrogen nuclei, $N_{\rm
  rec}=0.25$ is the mean number of recombinations per hydrogen nucleus
up to $z_{\rm re,full}$, and $\epsilon_{\rm AGN}$ is the AGN photon
emissivity per unit redshift, which is related to $\epsilon'_{\rm
  AGN}$ by $\epsilon_{\rm AGN}=\epsilon'_{\rm AGN}dt/dz$. The redshift
of at which reionization is $50\%$ complete, $z_{\rm re,half}$, is
calculated as in Eq(\ref{eq:ext_zre_calculation}), but for half the
threshold. To explore the effect of different assumptions for $f_{\rm
  esc}$, we allow this parameter to vary in the range $0-0.25$.

Fig.~\ref{fig:uncertain1} shows the effect of varying
  the AGN contribution (by varying $k_{\rm AGN}$) and $f_{\rm esc}$ on
  $z_{\rm re,half}$. We consider three models: \ Lacey16, SatFb and
  EvoFb, as indicated in the corresponding legends.  The contour lines
  show the predicted values of $z_{\rm re,half}$ in each model and
  the shaded area shows the region consistent with the \Planck data. The
  PL-EvoFb model is not considered here because it is disfavoured by
  the MW satellite metallicity data.

As we have seen, stars in the Lacey16 model do not
  produce enough ionizing photons to reionize the Universe
  sufficiently early; AGN can reionize the Universe in this model but
  only if their emissivity has a very flat slope, $-0.25\leq k_{\rm
    AGN}\leq -0.15$; this extreme region is illustrated in the upper
  left panel of Fig.~\ref{fig:uncertain1} as the red hatched area.
  The SatFb model also requires an AGN contribution in order to be
  consistent with the the values of $z_{\rm re,half}$ allowed by the
  \Planck data, but this is generally less than required for the
  Lacey16 model. For our fiducial value of $f_{\rm esc}=0.2$, the
  required AGN emissivity corresponds to $-0.49\leq k_{\rm AGN}\leq
  -0.15$; this region is the blue hatched area in the upper left panel
  of Fig.~\ref{fig:uncertain1}. For lower values of $f_{\rm esc}$, the
  required range of $k_{\rm AGN}$ shrinks and comes close to the
  allowed upper limit. Finally, the EvoFb model is consistent with the
  \Planck data in the case where all ionizing photons are produced by
  stars so long as $f_{\rm esc}\geq 0.07$; of course adding an AGN
  contribution makes it easier to reionize the Universe for even lower
  values of $f_{\rm esc}< 0.07$. 

In summary, even if AGN make a contribution to the
  ionizing photon budget, as long as $k_{\rm AGN}<-0.25$, our
  original, single power-law SN feedback model is incompatible with
  the \Planck data. If $k_{\rm AGN}<-0.49$
  and $f_{\rm esc}\geq 0.07$, then the evolving feedback model is
  preferred to the saturated feedback model, and our major conclusions
  regarding SN feedback still apply. Note that when $k_{\rm AGN}>-0.49$, the
  reionization redshift alone cannot discriminate between the SatFb and EvoFb
  models, but the measured far-UV galaxy luminosity functions at $z=7-10$
  (Fig.~\ref{fig:uv_lf}) still prefer the evolving feedback model.

\begin{figure} 
\centering
\includegraphics[width=7.0 cm]{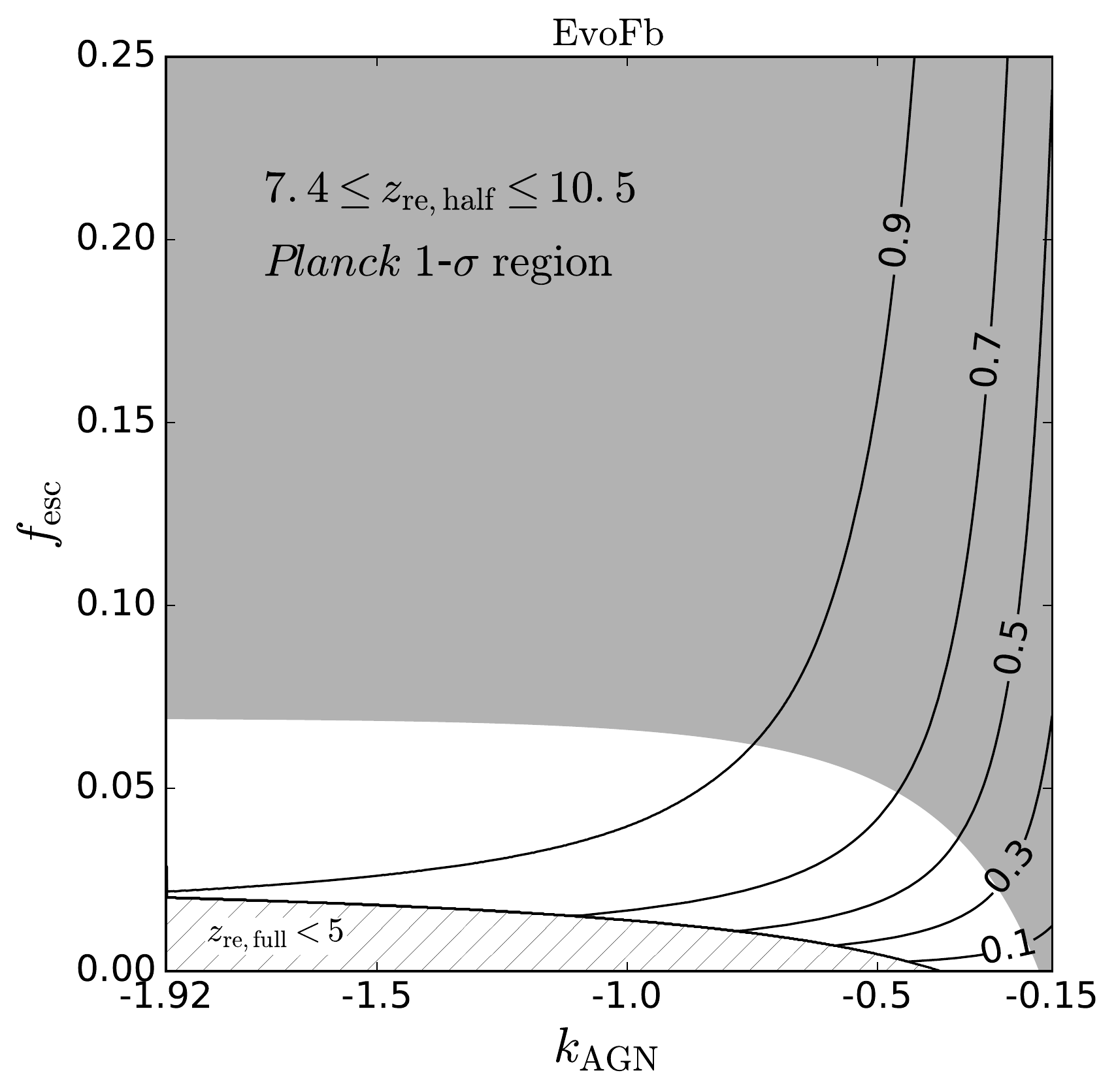}
\caption{The fraction of ionizing photons from stars
    for different combinations of $f_{\rm esc}$ and $k_{\rm AGN}$ for
    our best-fit model, EvoFb. This photon budget includes all 
  ionizing photons emitted from $z=\infty$ to $z_{\rm re,full}$. The fractions are shown as
    contour lines. The lower hatched region corresponds to $z_{\rm
      re,full} <5$ and is strongly excluded by other observations. The upper grey shaded
    region is allowed by the \Planck data.}
\label{fig:star_contribution}
\end{figure}

Our earlier conclusions regarding the sources of
  reionizing photons and their descendants are only valid when stars
  are the dominant source of reionizing photons.  The contour lines in
  Fig.~\ref{fig:star_contribution} show the fraction of the total
  ionizing photon budget produced by stars for different combinations
  of $f_{\rm esc}$ and $k_{\rm AGN}$. This photon budget includes all 
  ionizing photons emitted from $z=\infty$ to $z_{\rm re,full}$. This is only shown for the EvoFb
  model, because this is our best-fit model and thus the most relevant
  to a discussion of reionization sources and their descendants.  As
  the figure shows, so long as $f_{\rm esc}>0.07$ and $k_{\rm
    AGN}<-0.75$, over $90\%$ of the ionizing photons required for
  reionization come from stars; this fraction drops to $70\%$ if
  $k_{\rm AGN}=-0.49$, but is still dominant. Thus, our earlier
  conclusions regarding the reionization sources and their descendants
  remain valid so long as $f_{\rm esc}>0.07$ and $k_{\rm AGN}<-0.49$.

\section{summary} 
\label{sec:summary} 
We have investigated what constraints can be placed on supernova (SN)
feedback by combining a physical model of galaxy formation with
critical observations which constrain the strength of feedback in
opposite directions. The observational constraints are: the optical
and near-IR field luminosity functions (LFs) at $z=0$; the redshift
$z_{\rm re,half}$, at which the Universe was half reionized; the Milky
Way (MW) satellite LF; and the stellar metallicity vs. stellar mass
($Z_*\-- M_*$) relation for MW satellites.  We use the \GALFORM
semi-analytical model of galaxy formation embedded in the $\Lambda$CDM
model of structure formation, with 4 different
formulations for the mass-loading factor, $\beta$, of galactic
outflows driven by SN feedback: (a) in the Fiducial model, $\beta$ is
a simple power law in galaxy circular velocity, $\Vc$; (b) in the
Saturated feedback model, $\beta$ is a broken power law in $\Vc$, with
a flat slope at low $\Vc$; (c) in the power law Evolving
  feedback model, $\beta$ is a single power law in $\Vc$, but with a
  normalization that is lower at higher redshifts; (d) in the Evolving
  feedback model, $\beta$ decreases at high redshift, as well as
  having a break to a shallower slope at low $\Vc$. The Fiducial
model was previously tuned by \citet{Lacey16_model} to fit a wide
range of observational constraints, but not including reionization or
the MW satellites. Our main conclusions are:

\begin{enumerate}
\item The single power law formulation of $\beta$ as used in the
  Fiducial model can reproduce the faint ends of the $z=0$ field LFs
  and MW satellite LF, but leads to too low $z_{\rm re,half}$ and 
  too low MW satellite metallicities. This indicates that in this
  model, the SN feedback is too strong in small galaxies and/or at
  $z>8$.

\item Simply reducing the SN feedback in small galaxies, as in the
  Saturated model, does not provide an improvement relative to the
  single power law formulation of $\beta$.

\item The power law Evolving SN feedback model, with weaker
    SN feedback at high redshifts and stronger SN feedback at low
    redshifts, can successfully reproduce the faint ends of the $z=0$
    field LFs, $z_{\rm re,half}$ and the MW satellite LF, but still
    predicts MW satellite metallicities that are too low, indicating
    the necessity of weakening the SN feedback in low $\Vc$ galaxies.

\item The Evolving SN feedback model, with the SN
    feedback strength decreasing with increasing redshift and a
    saturation for $\Vc \leq 50\kms$ , seems to be preferred by the
  above mentioned observational constraints. Including the effects of
  local reionization may further improve the predictions for the MW
  satellite LF.

\item The physical reasons for the redshift evolution in our
  phenomenological Evolving SN feedback models could
  be that a single function of galaxy $\Vc$ only captures the effects
  of the gravitational potential well on the SN feedback, but the SN
  feedback is likely also to depend on factors such as the cold gas
  density and metallicity and the molecular gas fraction, which evolve
  with redshift. However, a more detailed ISM model is required to
  test the conclusions from this work further.

\item In all of the SN feedback models analysed in this work, around
  50\% of the photons which reionize the IGM are emitted by galaxies
  with stellar masses $M_*\gsim 10^9\Msol$, rest-frame far-UV absolute
  magnitudes, $M_{\rm AB}(1500{\rm \AA})\lsim -18$, galaxy circular
  velocities $\Vc \gsim 100\kms$ and halo masses $M_{\rm halo}\gsim
  10^{11}\Msol$ at the redshift $z\sim z_{\rm re,full}$ at which the
  Universe is fully reionized. In addition, most of the ionizing
  photons are predicted to be emitted by galaxies undergoing
  starbursts, rather than forming stars quiescently. This implies that
  the currently observed high redshift galaxy population should
  contribute about half of the ionizing photons that reionized
  Universe.

\item For our best fit model, namely the Evolving feedback model, the
  $z=0$ descendants of the major ionizing photon sources are
  relatively large galaxies with $M_{*}\gsim 10^{10}\Msol$, and are
  mainly in dark matter halos with $M_{\rm halo}\gsim
  10^{13}\Msol$. However, for these galaxies, the fraction of stars
  formed before reionization is low, while this fraction is high for
  dwarf galaxies with $z=0$ stellar masses $M_*<10^6\Msol$, even
  though the progenitors of such dwarfs contribute little to
  reionizing the Universe. This fraction also shows considerable
  scatter for the dwarfs, indicating that the star formation histories
  of these dwarf galaxies are very diverse.

\item For satellite galaxies in Milky Way-like halos, our best fit
  model implies that the fraction of stars formed before reionization
  is very low for large satellites like the LMC, SMC and Fornax, 
  but reaches very high values for very small satellites with stellar
  masses $M_*< 10^6\Msol$, like Leo II, Ursa Minor and Hercules, with
  median fractions around $80\%$, indicating that typically these small
  satellites formed most of their stars before reionization.

\end{enumerate}


\section*{Acknowledgements}
We thank Tom Theuns and Mahavir Sharma for helpful discussions. This
work was supported by the Science and Technology Facilities Council
grant ST/L00075X/1, and by European Research Council grant GA 267291
(Cosmiway), and SB is supported by STFC through grant ST/K501979/1. 
This work used the DiRAC Data Centric system at Durham
University, operated by the Institute for Computational Cosmology on
behalf of the STFC DiRAC HPC Facility (www.dirac.ac.uk). This
equipment was funded by BIS National E-infrastructure capital grant
ST/K00042X/1, STFC capital grant ST/H008519/1, and STFC DiRAC
Operations grant ST/K003267/1 and Durham University. DiRAC is part of
the National E-Infrastructure.

\bibliographystyle{mn2e}

\bibliography{paper}

\end{document}